\documentclass[conference]{IEEEtran}
\IEEEoverridecommandlockouts
\usepackage{amsmath,amssymb,amsfonts}
\usepackage{amsthm}
\usepackage{textcomp}
\usepackage{xcolor}
\usepackage{hyperref}  
\usepackage{url}  
\usepackage{booktabs}  
\usepackage{nicefrac}  
\usepackage{microtype} 
\usepackage{lipsum}
\usepackage{algorithm, algorithmic}
\usepackage{mathtools,nccmath}
\usepackage{array}
\usepackage{color}
\usepackage{fancyhdr}
\usepackage{graphicx}
\usepackage{epstopdf}

\def\BibTeX{{\rm B\kern-.05em{\sc i\kern-.025em b}\kern-.08em
    T\kern-.1667em\lower.7ex\hbox{E}\kern-.125emX}}
\begin{document}

\title{Conditional Analysis for Key-Value Data with Local Differential Privacy}


\author{\IEEEauthorblockN{Lin Sun}
    \IEEEauthorblockA{\textit{School of Software} \\
    \textit{Tsinghua University}\\
    Beijing, China \\
    sunl16@mails.tsinghua.edu.cn}
\and
    \IEEEauthorblockN{Jun Zhao}
    \IEEEauthorblockA{\textit{School of Computer Science and Engineering} \\
    \textit{Nanyang Technological University}\\
    Singapore\\
    junzhao@ntu.edu.sg}
\and
    \IEEEauthorblockN{Xiaojun Ye}
    \IEEEauthorblockA{\textit{School of Software} \\
    \textit{Tsinghua University}\\
    Beijing, China \\
    yexj@mail.tsinghua.edu.cn}
\and
    \IEEEauthorblockN{Shuo Feng}
    \IEEEauthorblockA{\textit{School of Software} \\
    \textit{Tsinghua University}\\
    Beijing, China \\
    fs15@mails.tsinghua.edu.cn}
\and
    \IEEEauthorblockN{Teng Wang}
    \IEEEauthorblockA{\textit{School of Computer Science and Technology} \\
    \textit{Xi'an Jiaotong University}\\
    Shaanxi, China \\
    wangteng0610@stu.xjtu.edu.cn}
\and
    \IEEEauthorblockN{Tao Bai}
    \IEEEauthorblockA{\textit{School of Computer Science and Engineering} \\
    \textit{Nanyang Technological University}\\
    Singapore\\
    bait0002@ntu.edu.sg}
}

\maketitle
\thispagestyle{plain} \pagestyle{plain}

\begin{abstract}
    Local differential privacy (LDP) has been deemed as the \textit{de facto} measure for privacy-preserving distributed data collection and analysis. Recently, researchers have extended LDP to the basic data type in NoSQL systems: the key-value data, and show its feasibilities in mean estimation and frequency estimation. In this paper, we develop a set of new perturbation mechanisms for key-value data collection and analysis under the strong model of local differential privacy. Since many modern machine learning tasks rely on the availability of conditional probability or the marginal statistics, we then propose the conditional frequency estimation method for key analysis and the conditional mean estimation for value analysis in key-value data. The released statistics with conditions can further be used in learning tasks. Extensive experiments of frequency and mean estimation on both synthetic and real-world datasets validate the effectiveness and accuracy of the proposed key-value perturbation mechanisms against the state-of-art competitors.
\end{abstract}


\newtheorem{theorem}{Theorem}
\newtheorem{corollary}{Corollary}[theorem]
\newtheorem{lemma}[theorem]{Lemma}
\newtheorem{definition}{Definition}

\section{Introduction}
    In the age of big data, personal-related data from user's side is routinely collected and analyzed by service providers to improve the quality of services. However, for the user side, directly sending original data can somehow lead to information leakage, which may draw potential privacy issues in many data-driven applications. To handle the privacy concerns, many mechanisms are proposed for privacy-preserving data analysis, among which stands out the differential privacy~\cite{dwork2006calibrating, dwork2014algorithmic, wang2018privtrie}.
    
    Usually, there are two kinds of differential privacy: the centralized setting and the local setting. In the centralized setting, the result of a query is computed, and then a noisy version of the output is returned (usually with Laplace noise~\cite{dwork2006calibrating}). In the local setting, the collecting and analyzing flow can be included into three steps: 1) Each record is first \textbf{encoded} into a specific data format (for example, by bloom filters). 2) Then the encoded data are \textbf{perturbed}. 3) At last, data from user side are \textbf{aggregated} and analyzed. Mechanisms with local differential privacy guarantee an individual's privacy against potential adversaries (including the aggregator in LDP). The local differential privacy has been widely used in crowdsourcing and IoT scenarios for privacy-preserving data analytics~\cite{To2018PrivacyPreservingOT, Fanaeepour2018histogramming}. To analyze user's data with high-level privacy guarantees, respected data service providers have applied local differential in their services. Google has proposed RAPPOR~\cite{erlingsson2014rappor} for crowdsourcing statistics in Chrome. Microsoft proposed a memoization mechanism for continual data collection. Apple has used differential privacy for frequency estimation~\cite{Apple2017ldp}, such as identifying popular emojis.
    
    Recently, a significant amount of attention has been focused on improving accuracy in mean and histogram estimation with local differential privacy guarantees, such as categorical values~\cite{erlingsson2014rappor}, set values~\cite{qin2016heavy, wang2018locally} and numerical values~\cite{NguyenSS16collect, Ning2019collecting}. For the first time, Ye~\textit{et al.}~\cite{qingqing2019privkv} formalize the frequency and mean estimation problems for key-value data under local differential privacy. Our work will improve previous studies in data collecting and analyzing of key-value data. For further clarification, we start from a dietary rating example.
    
    \begin{table}[t]
        \centering 
        \caption{Example for key-value data} 
        \label{tbl: consumer kv data}
        \begin{tabular}{|p{1cm}<{\centering}|p{6.5cm}<{\centering}|}  
            \hline  
            \textbf{Users} & \textbf{Food and ratings} \\
            \hline
            user 1 & $\langle \text{Hamburger}, 0.65 \rangle, \langle \text{Mash}, 0.8 \rangle, \langle \text{Pepsi}, 0.8 \rangle$ \\
            \hline
            user 2 & $\langle \text{Hamburger}, 0.8 \rangle, \langle \text{Fries}, -0.3 \rangle, \langle \text{Pepsi}, 0.9\rangle $ \\
            \hline
            user 3 & $\langle \text{Salad}, 0.75 \rangle, \langle \text{Fries}, -0.1 \rangle, \langle \text{Pepsi}, 0.8\rangle $ \\
            \hline
            ... & ... \\
            \hline
        \end{tabular}
        \vspace{-10pt}
    \end{table}
    
    \textbf{Motivating example}. As shown in Table~\ref{tbl: consumer kv data}, assume that analysts consider ratings of the food by collecting users' scores on each specific food. Each record consists of a set of key-value data and represents the orderings of an individual, where each key-value data shows one's appetite on the given food (note that the rating scores are normalized into $[-1,1]$). Based on the properties of key-value data, the analyzing tasks will include two parts: \textit{frequency estimation of keys} and \textit{mean estimation of values for given key}.
    
    The frequency estimation allows us to know the proportion of people with given keys. For example, the rate of people who ordered Hamburger can be given by $f_{k = \text{Hamburger}}$. The mean estimation allows us to know people's average appetite for the food they eat. For example, the average rating of $k = \text{Fries}$ is $\frac{-0.3 + (-0.1)}{2} = -0.2$, while that of $k = \text{Pepsi}$ is $\frac{0.8 + (-0.9) + 0.8}{3} = 0.83$, which might reveal to us that one who orders Pepsi loves it, but one orders Fries probably because he just needs something to eat.
    
    The \textit{PrivKV} and \textit{PrivKV}-based methods are proposed to handle the frequency and mean estimation in key-value data~\cite{qingqing2019privkv}. However, our experiments support that the \textit{PrivKV}-related methods only work well with a high average of mean. Thus in the first part of this paper, we propose a series of encoding and decoding mechanisms that are $\epsilon$-differentially private for frequency and mean estimation in key-value data. All the proposed methods have low communication cost and compared with \textit{PrivKVM}, and the proposed methods do not need to interact with an aggregator frequently.
    
    Our investigation indicates that correlations between attributes are essential in many analyzing and learning tasks. The On-Line Analytical Processing (OLAP) data cube is the exhaustion of possible marginals of data sets. And correlations are important in decision makings, such as the typical decision tree~\cite{deng2011bias}. For the key-value data, retrieving conditional probabilities between keys can provide useful information for more in-depth analysis.
    Thus the rest part of this paper focuses on analyzing correlations of frequencies and means between different keys. The problem is motivated by such kind of problems:
    
    \textbf{Motivating problem}: \textit{Will those ordered hamburgers order Pepsi? How to model people's appetites for Pepsi if they've ordered fries?} These questions are important to merchants. For example, upon knowing that people who love hamburger also wants to eat fries, a new combo can be introduced by merchants. Such problems are challenging when privacy concerns are considered. We call such kind of problems conditional analysis. 
    
    Currently, to the best of our knowledge, no proposed methods can handle the conditional analysis for key-value data. Even though the \textit{PrivKV}-based mechanisms can estimate frequency and mean for a single key, they do not support any conditional analysis as each user only randomly sends one key-value pair to the aggregator. To address these challenges, in the rest part of this paper, we introduce the conditional analysis mechanism for frequency and mean estimation. We define the $L$-way conditional notion and propose analyzing mechanism with $\epsilon$-local differential privacy guarantees. To summarize, the main contributions are listed as follows:
    
    \begin{itemize}
        \item We propose a new estimator for frequency and mean under the framework of \textit{PrivKV}.
        \item We propose several mechanisms for estimating the number of key-value data under the framework of LDP. Compared with existing algorithms, the proposed mechanisms are more effective and stable.
        \item For the first time, we introduce conditional analysis for key-value data. We formulate the problem of $L$-way conditional analysis in the local setting.
    \end{itemize}
    
    The rest of this paper is organized as follows. In Section~\ref{sec: Preliminaries}, we briefly include previous work. In Section~\ref{sec: Estimator for privkv} we propose a new estimator under the encoding results of \textit{PrivKV}. In Section~\ref{sec: Encoding Mechanisms for Key-Value Data}, several perturbation mechanisms are presented and analyzed. In Section~\ref{sec: Conditional frequency and means}, we define the conditional analysis problem and proposed methods for it. At last, the experimental results are shown in Section~\ref{sec: experiments} and the whole paper is concluded in Section~\ref{sec: conclusion}.

\section{Preliminaries}
\label{sec: Preliminaries}

\subsection{Local Differential Privacy}

    The notion of differential privacy was initially proposed for statistical database, where a trusted data curator is assumed. The data curator gathers, processes, and publishes data in a way that satisfies requirements of differential privacy. In many application scenarios, such a trusted party does not exist, therefore comes up the local differential privacy. 
    
    \begin{definition}[Local Differential Privacy~\cite{duchi2013local}]
        A randomized mechanism is $\epsilon$-local differential privacy ($\epsilon$-LDP) iff for any two tuples $x, x' \in \mathcal{X}$ and any output $o \in \mathcal{O}$: 
        
        \begin{equation}
            \Pr[\mathcal{M}(x) = o] \le e^\epsilon \cdot \Pr[\mathcal{M}(x') = o],\nonumber
        \end{equation}
        where the randomness is over mechanism $\mathcal{M}$.
    \end{definition}
    
    There are many functional properties of differential privacy, one of which is the composition~\cite{dwork2014algorithmic}, usually used to track the privacy loss in sequential executions.
    
    \begin{lemma}[Composition Theorem]
    \label{theorem: composition theorem}
        Let $\mathcal{M}_i: \mathbb{N}^{\lvert \mathcal{X}\rvert} \rightarrow \mathbb{R}_i$ be an $\epsilon_i$-differential privacy algorithm, and $\mathcal{M}(x): \mathbb{N}^{\lvert \mathcal{X}\rvert} \rightarrow \prod_{i=1}^k \mathbb{R}_i$, where $\mathcal{M}(x)$ is defined to be $\mathcal{M}(x) = \{\mathcal{M}_1(x), ..., \mathcal{M}_k(x)\}$. Then $\mathcal{M}(x)$ satisfies $\sum_{i=1}^k \epsilon_i$-differential privacy.
    \end{lemma}
    
    The canonical solution towards local differential privacy is randomized response~\cite{warner1965randomized}, first introduced in the literature as a survey design technique. The randomized response mechanism provides plausible deniability. Thus the aggregator cannot reveal original data with a high confidence level. The randomized response works as follows: for a user having a bit value $v \in \{0, 1\}$, he flips the value with rate $p > 0.5$ (which means the sent value is the same as the original value with probability $p$). Then the randomized response achieves LDP with $e^\epsilon = p / (1-p)$. Perturbation mechanisms with randomized response offer acceptable accuracy under large datasets. The randomized response plays a core role in many recent LDP mechanisms. One typical implementation is Google deployment of RAPPOR~\cite{erlingsson2014rappor}, the randomized response is used in each bit of output array by bloom filters. For continuous data, the randomized response can be used in the discretized value for mean estimation~\cite{duchi2014privacy, NguyenSS16collect, Duchi2018Minimax, wang2019locally}.
    
    The basic randomized response only works for binary response ($\lvert \mathcal{K} \rvert = 2$). For category data with $\lvert \mathcal{K} \rvert > 2$, a generalized randomized response (also called Direct Encoding~\cite{wang2017locally}) is proposed.
    
    \begin{equation}
        \Pr[\mathcal{M}(x) = v] = \begin{cases}
            p, &\text{if } v = x,\nonumber\\
            \frac{1-p}{\lvert \mathcal{K} \rvert-1}, &\text{otherwise.}
        \end{cases}
    \end{equation}
    
    In other words, the true value is reported with probability $p$, while each other value is reported with probability $\frac{1-p}{\lvert \mathcal{K} \rvert-1}$. To achieve $\epsilon$-LDP, we set $p = e^\epsilon / (e^\epsilon + \lvert \mathcal{K} \rvert - 1)$.

    
\subsection{Key-Value Data Collection}
    
    The problem of privacy-preserving key-value data collection with frequency and mean estimation in the local setting is first proposed in~\cite{qingqing2019privkv}.  Before defining the problem, we first describe the notations used in this paper. 
    
    \begin{table}[t]
    \centering 
    \caption{Notations} 
    \label{tab: notations} \vspace{-6pt}
    \begin{tabular}{|p{1.5cm}<{\centering}|p{5cm}<{\centering}|}
        \hline  
        \textbf{Symbol} & \textbf{Description} \\
        \hline
        $\mathcal{U}$ & the set of users, and $n= \lvert \mathcal{U} \rvert$\\
        \hline
        $u_{i:i\in [n]}$ & the $i$-th user in $\mathcal{U}$\\
        \hline
        $\mathcal{K}$ & the set of keys, and $d = \lvert \mathcal{K} \rvert $\\
        \hline
        $S_i$ & the set of key-value pair owned by $u_i$ \\
        \hline
        $\langle k_{i,j}, v_{i,j} \rangle$ & the $j$-th key-value pair in $S_i$ \\
        \hline
        $f_k$ & the frequency of key $k$ \\
        \hline
        $m_k$ & the mean of values with key $k$ \\
        \hline
        $\mathcal{C} = (\alpha, \beta)$ & conditions of keys \\
        \hline
    \end{tabular}
    \end{table}
    
    The key-value data collecting and analyzing framework under LDP can be briefly stated as follows: let the universe contain a set of users $\mathcal{U} = \{u_1, u_2, ..., u_n\}$ and a set of keys $\mathcal{K} = \{1, 2, ..., d\}$. The value domain $\mathcal{V} $ for the keys is in the domain of $[-1, 1]$. We consider that each user $u_i$ owns a list of (say $\ell_i$, which is at most $d$) key-value pairs $S_i : = \{ \langle k_{i,j},v_{i,j} \rangle \,\mid\, 1\leq j \leq \ell_i, k_{i,j}\in\mathcal{K}, v_{i,j}\in\mathcal{V}\}$. An untrusted data collector needs to estimate statistics information of the key-value data, especially, the frequency estimation and the mean estimation.
    
    \begin{itemize}
        \item \noindent \textbf{Frequency estimation:} The goal of frequency estimation is to estimate the frequency of key $k$. It is the portion of users who possess the key. It is defined as:
            \begin{equation}
                f_k = \frac{\lvert \{u_i | \exists \langle k,v\rangle \in S_i\} \rvert }{n} \nonumber.
            \end{equation}
        \item \textbf{Mean estimation:} The goal of mean estimation is to estimate the mean values of key $k$. It is defined as:
            \begin{equation}
                m_k = \frac{\sum_{i} \sum_{j:k_{i,j}=k} v_{i,j} }{ \lvert\{u_i | \exists \langle k,v\rangle \in S_i\}  \rvert} \nonumber.
            \end{equation}
    \end{itemize}
    
\subsection{\textit{PrivKV}}
    Ye~\textit{et al.}~\cite{qingqing2019privkv} adopt local perturbation protocol and propose the basic \textit{PrivKV} algorithm for statistical estimation. They also extend \textit{PrivKV} to \textit{PrivKVM} and \textit{PrivKVM}$^+$ to improve estimation accuracy.
    
    If the $j$-th key $k_{i,j}$ of user $i$ exists, we have $k_{i,j}= 1$, otherwise, $k_{i,j} = 0$. When the key does not exist in $u_i$, it is represented by $\langle k,v\rangle = \langle 0,0\rangle$. Thus, given $k$, the key-value pair can be represented as $\langle k,v\rangle \in \{\langle 0,0\rangle, \langle 1, v\rangle\}$ where $v \in [-1, 1]$. To achieve LDP in key-value protection, the \textit{PrivKV}-based mechanisms have four types of perturbations:
    
    \begin{itemize}
        \item $1 \rightarrow 1$: The key exists before and after perturbation. Under the circumstances, only the value needs to be perturbed, i.e., $\langle 1, v\rangle \rightarrow \langle 1, v'\rangle$.
        \item $1 \rightarrow 0$: The key-value pair disappears after perturbation. As the key does not exist, the value is meaningless and is set to zero. For example, $\langle 1,v \rangle \rightarrow \langle 0,0 \rangle$.
        \item $0 \rightarrow 1$: A new key-value pair is generated after perturbation and a value is assigned, i.e., $\langle 0, 0 \rangle \rightarrow \langle 1, v'\rangle$.
        \item $0 \rightarrow 0$: The key-value pair does not exist before and after the perturbation. In this case, the key-value pair is kept unchanged, i.e., $\langle 0,0\rangle \rightarrow \langle 0,0\rangle$.
    \end{itemize}
    
    The \textit{PrivKV}-based mechanisms guarantee $\epsilon$-LDP by providing indistinguishability for both key and value in key-value data. The randomized response can be directly used for key perturbation, as the key space is binary. For value perturbation, the perturbation mechanism called \textit{Harmony}~\cite{NguyenSS16collect} is used for mean estimation (also seen in~\cite{ding2018comparing}). Values in continuity interval $[-1,1]$ are first discretized to $\{-1,1\}$ through Eq.~(\ref{equ: discretization}), then the randomized response is used in the discretized value for perturbation. These two steps are called VPP (\textbf{V}alue \textbf{P}erturbation \textbf{P}rimitive).
    

    \begin{align}
    \label{equ: discretization}
        v^* = \begin{cases}
        1 & \quad  \text{with probability } \frac{1+v}{2}, \\
        -1 & \quad \text{with probability } \frac{1-v}{2}.
        \end{cases}
    \end{align}

    \begin{algorithm}[t] 
    \caption{Local Perturbation Protocol (LPP)} 
    \label{alg: local perturbation protocol} 
    \begin{algorithmic}[1] 
        \REQUIRE ~~\\ 
        User $u_i$'s set of key-value pairs $S_i$; Privacy budget $\epsilon_1$ and $\epsilon_2$.
        \ENSURE ~~\\ 
        $LPP(S_i, \epsilon_1, \epsilon_2$) is the perturbed key-value pair $\langle k_j, v^*\rangle$ of the $j$-th key.
        \STATE Sample $j$ uniformly at random from $[d]$;
        \IF{$k_j$ exists in the key set of $S_i$} 
            \STATE $V^* = VPP(V_j, \epsilon_2)$;
            \STATE Perturbs $\langle k_j, v^*\rangle$ as:
            \begin{equation}
                \langle k_j, v^* \rangle = \left\{ \begin{array}{ll}
                \langle 1, v^*\rangle & \quad  \text{with probability } \frac{e^{\epsilon_1} }{e^{\epsilon_1}+1},\nonumber\\
                \langle 0, 0\rangle & \quad \text{with probability } \frac{1}{e^{\epsilon_1}+1};\\
                \end{array} \right.
            \end{equation}
        \ELSE
            \STATE Randomly draw a value $m\in [-1,1]$;
            \STATE $v^* = VPP(m, \epsilon_2)$;
            \STATE Perturbs $\langle k_j, v^*\rangle$ as:
            \begin{equation}
                \langle k_j, v^*\rangle = \left\{ \begin{array}{ll}
                \langle 0, 0\rangle & \quad  \text{with probability } \frac{e^{\epsilon_1} }{e^{\epsilon_1}+1},\nonumber\\
                \langle 1, v^*\rangle & \quad \text{with probability } \frac{1}{e^{\epsilon_1}+1};\\
                \end{array} \right.
            \end{equation}
        \ENDIF
        \RETURN $j$ and $\langle k_j, v^*\rangle$; 
    \end{algorithmic}
    \end{algorithm}
    
    By assigning the randomized response to the key and the value perturbation mechanism to the value, the local perturbation protocol for key-value data is drawn (Algorithm~\ref{alg: local perturbation protocol}). When an untrusted aggregator receives the perturbed key-value data, he can then estimate the frequency of keys and mean of values with the \textit{PrivKV} algorithms (Algorithm~\ref{alg: PrivKV}). Based on the \textit{PrivKV} algorithm, the \textit{PrivKVM} with iterations and \textit{PrivKVM}$^+$ with virtual iterations are also proposed. For simplicity, the algorithms are not detailed. The results are covered in our experimental analysis.
    
    \begin{algorithm}[t] 
    \caption{\textit{PrivKV}} 
    \label{alg: PrivKV} 
    \begin{algorithmic}[1] 
        \REQUIRE ~~\\ 
        Key-value pairs $S={S_1, S_2, ..., S_n}$; Privacy budgets $\epsilon_1, \epsilon_2$.
        \ENSURE ~~\\ 
        Frequency vector $\boldsymbol{f}^*$ and mean vector $\boldsymbol{m}^*$.
        
        \STATE //\textbf{User-side perturbation}\COMMENT {\textbf{User-side perturbation}}
        \STATE Each user perturbs her set and sends the index $j$ and $\langle k_j, v^*\rangle = LPP(S_i, \epsilon_1, \epsilon_2)$ to data collector.
        \STATE //\textbf{Aggregator-side calibration}
        \FOR {each key $k$}
            \STATE Aggregator calculates frequency $f_k^*$;
            \STATE Aggregator calibrates the frequency as:
                \begin{equation}
                    f_k^*=\frac{p-1+f_k^*}{2p-1}, \text{where } p = \frac{e^{\epsilon_1}}{e^{\epsilon_1}+1}; \nonumber
                \end{equation}
                
            \STATE Aggregator counts 1 and -1 in the set of values:
                \begin{equation}
                    n_1' = Count(1), n_2' = Count(-1); \nonumber
                \end{equation}
                
            \STATE Let $N = n_1' + n_2'$.
            \STATE Aggregator calibrates the counts as ($p = \frac{e^{\epsilon_2}}{e^{\epsilon_2}+1}$):
                \begin{equation}
                    n_1^*= \frac{p-1}{2p-1}\cdot N + \frac{n_1'}{2p-1}, n_2^*= \frac{p-1}{2p-1}\cdot N + \frac{n_2'}{2p-1}; \nonumber
                \end{equation}
                
            \STATE Clip $n_1^*$ and $n_2^*$ to $[0, N]$;
            \STATE Aggregator calculates mean $m_k^* = \frac{n_1^*-n_1^*}{N}$;
        \ENDFOR
        \RETURN $\boldsymbol{f}^*$ and $\boldsymbol{m}^*$; 
    \end{algorithmic}
    \end{algorithm}
    
\section{Estimator for \textit{PrivKV}}
\label{sec: Estimator for privkv}

    The main intuition for the mean estimation under local differential privacy is to estimate the frequency of $k=1$. This problem has been well studied under current frequency estimation framework. After discretization, the number of $\langle k, v \rangle$ with $k=1$ consists of two parts: those with value $-1$ and those with value $1$. For the mean estimation, the top priority task is to estimate the number of key-value pair with value $1$ and $-1$, which is $m_k = \frac{N_{1} - N_{-1}}{N_{1}+N_{-1}}$. The perturbed key-value data are in the same space as that after discretization (i.e., all in $\{\langle 0, 0 \rangle, \langle 1, 1 \rangle, \langle 1, -1 \rangle\}$). The aggregator only uses the counting information of $\{\langle 1, 1 \rangle, \langle 1, -1 \rangle\}$ according to the \textit{PrivKV} algorithm (note that the $\{\langle 1, 1 \rangle, \langle 1, -1 \rangle\}$ is the perturbed value). This causes error for the mean estimation (Line 7-8 in Algorithm~\ref{alg: PrivKV}), as part of $\langle 0, 0 \rangle$ also turns into $\{\langle 1, 1 \rangle, \langle 1, -1 \rangle\}$ and some key-value pairs turns into $\langle 0, 0 \rangle$.
    
    It inspires us to develop mean estimation method to eliminate the impact of key perturbation. Instead of directly estimating $m_k$ with received key-value pairs, we design an unbiased estimator for estimating $N_{0}, N_{1}$ and $N_{-1}$.
    
    For the aggregator, let $M_{1} = Count(\langle 1, 1\rangle), M_{-1} = Count(\langle 1, -1\rangle)$ be the counts of the key-value pairs $\langle 1, 1\rangle, \langle 1, -1\rangle$ respectively, and $M_{0} = Count(\langle 0, 0 \rangle)$ be the counts of the received key-value pairs without key. Then the total received records by the data collector is $M = M_{0} + M_{1} + M_{-1}$. Let $p_1' = 2p_1 - 1$ and $p_2' = 2p_2-1$, according to the encoding process, we can estimate $N_{1}$ and $N_{-1}$ by $N_{1}^*$ and $N_{-1}^*$:
    
    
    
    \begin{equation}
    \label{equ: add1}
        N_1^*= \frac{(p_1 p_2' + p_1')M_1 + (p_1 p_2' - p_1')M_{-1} - p_1 p_2' (1-p_1) M}{2p_1 \cdot p_1' \cdot p_2'}, \nonumber
    \end{equation}
    
    \begin{equation}
    \label{equ: sub1}
        N_{-1}^* = \frac{(p_1 p_2' - p_1')M_1 + (p_1 p_2' + p_1')M_{-1} - p_1 p_2' (1-p_1) M}{2p_1 \cdot p_1' \cdot p_2'}. \nonumber
    \end{equation}
    
    \begin{theorem}
        The estimators of $N_{1}^*$ and $N_{-1}^*$ for $N_{1}$ and $N_{-1}$ are unbiased, respectively.
        \begin{proof}
            Instead of retrieving $\mathbb{E}[N_1^*] = N_1$ and $\mathbb{E}[N_{-1}^*] = N_{-1}$, we calculate by transforming the above equations. Through the encoding process of Algorithm~\ref{alg: local perturbation protocol}, we have:
            
            \begin{equation}
                \begin{cases}
                    M_{0}= N_{0} \cdot p_1 + (N_{1}+N_{-1}) \cdot (1-p_1), \nonumber\\
                    M_{1}= N_{0} \cdot \frac{1-p_1}{2} + N_{1} \cdot p_1 \cdot p_2 + N_{-1} \cdot p_1 (1-p_2), \nonumber\\
                    M_{-1} = N_{0} \cdot \frac{1-p_1}{2} + N_{1} \cdot p_1 \cdot (1-p_2) + N_{-1} \cdot p_1 \cdot p_2.\nonumber
                \end{cases}
            \end{equation}
            
            From which we get:
            
            \begin{equation}
                \begin{cases}
                    \mathbb{E}[N_1^*] + \mathbb{E}[N_{-1}^*]= \frac{M_1 + M_{-1} - M(1-p_1)}{2p_1-1} = N_1 + N_{-1},\nonumber\\
                    \mathbb{E}[N_1^*] - \mathbb{E}[N_{-1}^*]= \frac{M_1 - M_{-1}}{p_1 (2p_2 - 1)} = N_1 - N_{-1}.
                \end{cases}
            \end{equation}
            
            Then it holds that:
            
            \begin{equation}
                \begin{cases}
                    \mathbb{E}[N_1^*]= N_1, \\
                    \mathbb{E}[N_{-1}^*]= N_{-1}. \nonumber
                \end{cases}
            \end{equation}
            which concludes that the $N_1^*$ and $N_{-1}^*$ are unbiased estimator for $N_1$ and $N_{-1}$. We can also estimate $N_0$ by $N_0^*= M - N_1^* - N_{-1}^*$, which is also unbiased.
        \end{proof}
    \end{theorem}

\section{LDP for Key-Value Data}
\label{sec: Encoding Mechanisms for Key-Value Data}

    In this section, we combine the state-of-art locally differentially private mechanisms for data collecting and propose several $\epsilon$-LDP perturbation mechanisms for key-value data collecting and analyzing that can be used in different scenarios.
    
\subsection{\textit{F2M}: Frequency to Mean}
    
    Unlike \textit{PrivKV}-based mechanisms, we notice that there is no need to maintain the authenticity of the sent key-value pairs. For example, when original key $k_i$ does not exist in key-value pairs, the data should be in form of $\langle k_i, v_i\rangle = \langle 0, 0 \rangle$. Setting the value $v_i$ to any value will make it meaningless. Thus, in the $PirvKV$ algorithm, the perturbed key-value results can only be in the form of $\langle 0,0\rangle$, $\langle 1, -1\rangle$ or $\langle 1, 1\rangle$, where $\langle k_i, v_i\rangle = \langle 0,0\rangle$ represents that the key does not exists and $v_i=0$ is useless. Whereas we think more states in the perturbed space can provide more information when estimating. From this point, the state $\langle 0, 0 \rangle$ is substituted by $\langle 0, 1 \rangle$ and $\langle 0, -1 \rangle$. 
    
    \begin{algorithm}[t] 
    \caption{\textit{F2M}: Frequency to Mean} 
    \label{alg: KV perturbation}
    \begin{algorithmic}[1] 
        \REQUIRE ~~\\ 
        User $u_i$'s set of key-value pairs $S_i$; Privacy budget $\epsilon_1$ and $\epsilon_2$; Default value $\overline{v}$ when key does not exist.\\
        \ENSURE ~~\\ 
        $F2M(S_i, \epsilon_1, \epsilon_2$) is the perturbed key-value pair
        $\langle k_{i,j}', v_{i,j}'\rangle$ of the $j$-th key of user $i$
        \STATE Sample $j$ uniformly at random from $[d]$;
        \STATE Perturb key with:
            \begin{equation}
                \Pr[k_{i,j}' =k_{i,j}] = \frac{e^{\epsilon_1}}{e^{\epsilon_1}+1};\nonumber
            \end{equation}
            
        \IF{$k_j$ exists in the key set of $S_i$} 
            \STATE $v_{i,j}' = VPP(v_{i,j}, \epsilon_2)$;
        \ELSE
            \STATE $v_{i,j}' = VPP(\overline{v}, \epsilon_2)$;
        \ENDIF
        \RETURN $j$ and $\langle k_j, v^* \rangle$; 
    \end{algorithmic}
    \end{algorithm}

    In the LPP, when an existing key of a key-value data is perturbed to not exist, the value is directly set to 0, which increases error in mean estimation. The main difference between Algorithm~\ref{alg: KV perturbation} and Algorithm~\ref{alg: local perturbation protocol} is that all the outputs of LPP are in the same space as the original space, meanwhile $F2M$ of Algorithm~\ref{alg: KV perturbation} is not. In this mechanism, we treat key and value as irrelevant data and perturb them separately. 
    
    Perturbing key and value of a key-value pair independently allows the aggregator to estimate the frequency and mean of all key-value pairs. However, the goal of mean estimation in key-value data is to estimate those values with keys. Hence the influence of frequency to mean estimation should be considered. When encoding, we set the value without key to a default value $\overline{v}$. We will further explain how to use this information for mean estimation. Let:
    
    \begin{equation}
        \begin{cases}
            M_{-1}  &= Count(\langle *, -1 \rangle),\nonumber\\
            M_1 &= Count(\langle *, 1 \rangle).
        \end{cases}
    \end{equation}
    
    Then we can estimate the mean of all the values (with and without key) by:
    
    \begin{equation}
        m^*_{all} = \frac{e^\epsilon+1}{e^\epsilon-1} \cdot \frac{M_1 - M_{-1}}{M_1 + M_{-1}}.\nonumber
    \end{equation}
    
    With $m^*_{all}$ and $f^*$, we can then estimate $m_k^*$ by:
    
    \begin{equation}
        m_k^*=\frac{m^*_{all} - (1-f_k^*)\cdot \overline{v}}{f_k^*}.
    \end{equation}
    
   In the \textit{F2M}   mechanism of Algorithm~\ref{alg: KV perturbation}, we perturb key and value separately. From the  composition theorem (Lemma~\ref{theorem: composition theorem}),  \textit{F2M} achieves $\epsilon$-local differential privacy. Also, we can be sure with probability at least $(1-\delta)^2$ that (in Appendix~\ref{appendix: f2m}):
    
    \begin{equation}
        |m_k^* - m_k| \le \frac{2(f_k+1)(e^\epsilon+1)\sqrt{\ln{(2/\delta)}}}{\sqrt{2N}f_k^2 (e^\epsilon-1) - f(e^\epsilon-1)\sqrt{\ln{(2/\delta)}}} .\nonumber
    \end{equation}

\subsection{Unary Encoding for Key-Value data}
    
    The \textit{F2M} perturbation mechanism aims at eliminating the restriction that the value can only be 0 when the key of a key-value pair does not exist. Thinking that the original key-value data can only be in three statures when discretized, we pool the principle of the generalized randomized response to design a mapping function between the original and perturbed space. 
    
    For $\langle k',v'\rangle \in \{\langle 0, 0 \rangle, \langle 1, 1\rangle, \langle 1, -1\rangle\}$, the mapping between perturbed key-value data and the original data is designed by: 
    \begin{equation}
    \label{equ: unary encoding}
        \Pr[\langle k', v' \rangle ] = \frac{e^\epsilon-1}{e^\epsilon+2}\cdot [(k' \odot k) \land (v' \odot v^*)] + \frac{1}{e^\epsilon+2}.
    \end{equation}
    where $v^* = \text{Discretiaztion}(v)$ represents the discretization process shown in Eq.~(\ref{equ: discretization}), and $(k' \odot k) \land (v' \odot v^*)$ equals $1$ if $\langle k', v' \rangle = \langle k, v^* 
        \rangle$ and $0$ otherwise. We name this perturbation mechanism \textit{KVUE} (\textbf{K}ey-\textbf{V}alue \textbf{U}nary \textbf{E}ncoding). The probability mapping function is somehow difficult to understand. If we treat each key-value pair as a whole entity instead of treating key and value separately, we can directly use the generalized randomized response. Intuitively inspired by this, the mapping is equal to: 
    \begin{equation}
        \Pr[\langle k', v' \rangle = \langle k, v^* 
        \rangle] = \frac{e^\epsilon}{e^\epsilon+2}\nonumber
    \end{equation}
    
    \begin{algorithm}[t] 
    \caption{\textit{KVUE}: Unary Encoding for Key-Value Data} 
    \label{alg: KV Unary Encoding}
    \begin{algorithmic}[1] 
        \REQUIRE ~~\\ 
        User $u_i$'s set of KV pairs $S_i$; Privacy budget $\epsilon$. \\
        \ENSURE ~~\\ 
        $KVUE(S_i, \epsilon$) is the perturbed KV pair
        $\langle k_j, v^*\rangle$ of the $j$-th key
        \STATE Sample $j$ uniformly at random from $[d]$.
        \IF{$k_j$ exists}
            \STATE Discretization $v$ to $v^*$.
        \ENDIF 
        \STATE Report $\langle k',v'\rangle \in \{\langle 0, 0 \rangle, \langle 1, 1\rangle, \langle 1, -1\rangle\}$ with Eq.~(\ref{equ: unary encoding}).
        \RETURN $j$ and $\langle k', v'\rangle$; 
    \end{algorithmic}
    \end{algorithm}
    
    \begin{theorem}
        The unary encoding mechanism for key-value pair achieves $\epsilon$-LDP.
        \begin{proof}
        According to Eq.~(\ref{equ: unary encoding}), for any key-value pairs $\langle k, v \rangle \in \{\langle 0,0 \rangle, \langle 1, v \rangle\}$ and the possible output $\langle k',v' \rangle$, we have:
        \begin{equation}
            \frac{1}{e^\epsilon+2} \le \Pr[\langle k', v' \rangle | \langle k, v \rangle] \le \frac{e^\epsilon}{e^\epsilon+2}.\nonumber
        \end{equation}
        
        Thus, for any input $\langle k_1, v_1 \rangle, \langle k_2, v_2 \rangle$ and output $\langle k_o, v_o \rangle$, we obtain $\Pr[\langle k_o, v_o \rangle | \langle k_1, v_1 \rangle] \le e^\epsilon \times \Pr[\langle k_o, v_o \rangle | \langle k_2, v_2 \rangle], $
       which ensures $\epsilon$-LDP for $KVUE$.
        \end{proof}
    \end{theorem}
    
    Same as the analysing in \textit{F2M}, let $M_{0}, M_{1}, M_{-1}$ be the number of $\langle 0,0 \rangle$, $\langle 1, 1 \rangle$, $\langle 1, -1 \rangle$ received by the aggregator. For $i \in \{0, 1, -1\}$, denote (let $p = \frac{e^\epsilon}{e^\epsilon + 2}$):
    
    \begin{equation}
        N_{i}^* = \frac{2M_i - (1-p)M}{3p-1}.
    \end{equation}
    
    Then we can say that $N_{i}^*$ is an unbiased estimator. The proof is given in Theorem~\ref{theorem: unbiased of unary encoding}.
    
    \begin{theorem}
    \label{theorem: unbiased of unary encoding}
        $\forall i \in \{0, 1, -1\}$, $N_{i}^*$ is an unbiased estimator.
        \begin{proof}
            According to the algorithm, for $i,j \in \{0, 1, -1\}$, we have:
            \begin{equation}
            M_i =p  N_i + \sum_{j\not= i }   \frac{1-p}{2} N_j.\nonumber
            \end{equation}
            
            We then achieve:
            
            \begin{align}
                \mathbb{E}[N_i^*] & = \frac{2\mathbb{E}[M_i] - (1-p)M}{3p-1} \nonumber\\
                & = \frac{2N_i p + (1-p)\sum_{j\not= i}N_j - (1-p)M}{3p-1} \nonumber\\
                & = N_i.\nonumber
            \end{align}
            
            which concludes the correctness of theorem.
        \end{proof}
    \end{theorem}
    
    The unary encoding first maps a key-value pair into a single item and then uses the generalized randomized response to achieve $\epsilon$-LDP. Thus the variance is the same as the that of Direct Encoding (DE~\cite{wang2017locally}):
    \begin{equation}
        Var(N_i) = \frac{N \cdot (e^\epsilon+1)}{(e^\epsilon-1)^2}.
    \end{equation}
    
    \begin{theorem}\label{theorem-5}
        Given $\delta \in (0,1)$, with probability at least $1-\delta$, we have:
        \begin{equation}
            \left | N_i^*- N_i \right |\leq \frac{e^\epsilon +2}{e^\epsilon -1} \sqrt{\frac{N}{2}\cdot \ln \frac{2}{\delta} }. \nonumber
        \end{equation}

        \begin{proof}
            Based on the Chernoff--Hoeffding bound~\cite{ding2017collecting}, for every $t>0$, it holds that:
            \begin{equation}
                \Pr\big[\lvert M_i^* - M_i \rvert \ge t\big] \le 2 e^{- \frac{2t^2}{N}}.\nonumber
            \end{equation}
            
            Then, we have:
            \begin{equation}
                \Pr\big[\lvert N_i^* - N_i\rvert \ge \frac{2t}{3p-1}\big] \le 2e^{- \frac{2t^2}{N}}.\nonumber
            \end{equation}
            
            Let $r = \frac{2t}{3p-1}$, which corresponds to $t = r(3p-1)/2$, we achieve:
            
            \begin{equation}
                \Pr[\lvert N_i^* - N_i \rvert \ge r] \le 2e^{-\frac{r^2 (3p-1)^2}{2N}}.\nonumber
            \end{equation}
            
            Let $\delta = 2e^{-\frac{r^2(3p-1)^2}{2N}}$, corresponding to $r = \frac{1}{3p-1} \sqrt{ 2N\cdot \ln \frac{2}{\delta} }$, then we can say that with probability at least $1-\delta$, we have:
            
            
            \begin{equation}
                \lvert N_i^* - N_i \rvert  \le \frac{e^\epsilon +2}{e^\epsilon -1} \sqrt{\frac{N}{2}\cdot \ln \frac{2}{\delta} }.
            \end{equation}
            
            Thus, this completes the proofs.
        \end{proof}
    \end{theorem}
    
    With the unbiased estimator for $N_0, N_1$ and $N_{-1}$, we can then estimate the frequency and mean easily.

\subsection{One-Hot Encoding for Key-Value Data}
    
    The one-hot encoding mechanism was commonly used in histogram estimation~\cite{yiwen2019utility, erlingsson2014rappor}. For bucket $i$, the one hot encoding returns a vector in which all bits are $0$ except the $i$-th index. The encoding mechanism inspires us to uses such a bit array for state representation of key-value data. As analyzed in preceding sections, there are three statuses when one key-value pair is discretized. We design the following mechanism for index projection:
    
    \begin{equation}
        \mathcal{I}(\langle k, v \rangle) = k\cdot \text{Discretization}(v) + 1.\nonumber
    \end{equation}
    
    Thus the one-hot encoding for a key-value pair can be represented by $\mathcal{A}[\mathcal{I}(k,v)] = 1$. Then we use randomized response in each bit of $\mathcal{A}$ by:
    
    \begin{equation}\label{prob-kvoh}
        \Pr\big[\mathcal{A}[i] = 1\big] = \frac{e^{\epsilon/2}-1}{e^{\epsilon/2}+1}\cdot \mathcal{A}[i] + \frac{1}{e^{\epsilon/2}+1}.
    \end{equation}
    
   The randomized response guarantees $\epsilon/2$-LDP for every single bit. Since $A$ and $A'$ differ only in two bits, so we achieve $\epsilon$-LDP for $\mathcal{A}$ according to the composition theorem. We call this \textit{KVOH} (\textbf{K}ey-\textbf{V}alue \textbf{O}ne-\textbf{H}ot mechanism).
    
    \begin{algorithm}[t] 
    \caption{\textit{KVOH}: One Hot encoding for Key-Value data} 
    \label{alg: KV onehot}
    \begin{algorithmic}[1] 
        \REQUIRE ~~\\
        
            A user $u$'s set of key-value pairs $S$; Privacy budget $\epsilon$. \\
        \ENSURE ~~\\ 
            $KVOH(S, \epsilon$) is the perturbed KV pair.
        \STATE Sample $j$ uniformly at random from $[d]$.
        \IF {$k_j$ exists}
        \STATE Discretization $v$ to $v^*$.
        \ENDIF 
        \STATE Initialize a empty array: $\mathcal{A} = [0, 0, 0]$.
        \STATE Indexing: $\mathcal{A}[k\cdot v^* + 1] = 1$.
        \label{alg-line: indexing}
        \STATE Report each bit in $\mathcal{A}$ with probability defined in Eq.~(\ref{prob-kvoh}).
        \RETURN $\mathcal{A}$; 
    \end{algorithmic}
    \end{algorithm}

    
    Same as the proposed methods, we focus on estimating the number of each state after discretization instead of directly estimating frequency and mean, which is to retrieve the number of states in $\mathcal{A}$. Let $M_i = \sum \mathcal{A}_i$ be the sum of received arrays and $N$ denotes the number of arrays. We first adjust the sum of arrays before perturbation by:
    
    \begin{equation}
        N_i^*= \frac{(e^{\epsilon/2}+1)\cdot M_i - N}{e^{\epsilon/2}-1}. \nonumber
    \end{equation}
    
    It is easy to prove that the estimator of $N_i$ is unbiased \cite{wang2017locally}, and the variance is:
    \begin{equation}
        Var(N_i) = \frac{N\cdot e^{\epsilon/2}}{(e^{\epsilon/2}-1)^2}. \nonumber
    \end{equation}

    Similar to the proof of Theorem~\ref{theorem-5}, we can obtain
    \begin{equation}
        \Pr[|N_i^* - N_i| \ge t \cdot \frac{e^{\epsilon/2}-1}{e^{\epsilon/2}+1}] \le 2e^{-\frac{2t^2}{N}}. \nonumber
    \end{equation}
    
    By setting $r = t\cdot\frac{e^{\epsilon/2}-1}{e^{\epsilon/2}+1} $, we then have:
    
    \begin{equation}
        \Pr\big[|N_i^* - N_i| \ge r \big]\le 2e^{-\frac{2r^2}{N} \cdot (\frac{e^{\epsilon/2}-1}{e^{\epsilon/2}+1})^2}. \nonumber
    \end{equation}
    
    By using $\delta = 2e^{-\frac{2r^2}{N} \cdot (\frac{e^{\epsilon/2}-1}{e^{\epsilon/2}+1})^2}$, we can say that with probability at least $1-\delta$, we have:
    
    \begin{equation}
        |N_i^*-N_i| \le \frac{e^{\epsilon/2}+1}{e^{\epsilon/2}-1} \cdot \sqrt{\frac{N}{2} \cdot \ln \frac{2}{\delta}}. \nonumber
    \end{equation}
    
\subsection{Estimation Analysis}

    In the designation of \textit{KVOH} and \textit{KVUE} encoding mechanisms, we propose unbiased estimator for the states of $\langle 0,0 \rangle$, $\langle 1, -1 \rangle, \langle 1, 1\rangle$ (denoted as $N_0, N_A$ and $N_B$) and use the number of states for further estimation instead of directly estimating the frequency and mean:
    
    \begin{equation}
        f_k^*= \frac{N_A^* + N_B^*}{N},~\text{and}~ \text{\quad}m_k^*= \frac{N_A^*-N_B^*}{N_A^*+N_B^*}. \nonumber
    \end{equation}
    
    In this section, we give the upper bound for the estimator of $f^*$ and $m^*$. To analyze the estimation error, we first define $ \theta_A = N_A^* -N_A$ and $ \theta_B= N_B^* -N_B$ as the estimation error of the state number for different states. For the frequency estimation, we then can analyze the estimating error by:
    
    \begin{align}
        err(f_k) &=  \left |\frac{N_A^*+ N_B^*}{N}- \frac{N_A + N_B}{N} \right |\nonumber\\
        &= \left |\frac{\theta_A + \theta_B}{N} \right |.\nonumber
    \end{align}

    For mean estimation, we have:
    
    \begin{small}
    \begin{align}
        err(m_k) &= \left |\frac{N_A^*-N_B^*}{N_A^*+N_B^*} - \frac{N_A-N_B}{N_A+N_B} \right | \nonumber\\
        &= \left | \frac{N_A-N_B + \theta_A - \theta_B}{N_A+N_B + \theta_A + \theta_B} - \frac{N_A-N_B}{N_A+N_B} \right|\nonumber\\
        &= \left | \frac{(\theta_A - \theta_B) (N_A+N_B) - (\theta_A + \theta_B) (N_A-N_B)}{(N_A + N_B)(N_A + N_B + \theta_A + \theta_B)}\right| \nonumber\\
        &= \left | \frac{(\theta_A - \theta_B) - (\theta_A + \theta_B) \cdot m_k}{f_k \cdot N + \theta_A + \theta_B} \right|.\nonumber
    \end{align}
    \end{small}
    
    In the prior sections, we've proven that with probability at least $1-\delta$, we have $\Pr[|N_i^*-N_i|\le r]$. Considering the noises on $N_A$ and $N_B$ are independent, the error of frequency estimation can be sure with:
    \begin{equation}
        \Pr[err(f_k) \le \frac{2r}{N} ] \ge (1-\delta)^2.\nonumber
    \end{equation}
    
    For the mean estimation, we then have:
    \begin{align}
        err(m_k) &= \left| \frac{(\theta_A - \theta_B) - (\theta_A + \theta_B) \cdot m_k}{f_k \cdot N + \theta_A + \theta_B} \right| \nonumber\\
        &\le \left|\frac{\theta_A (1-m_k) - \theta_B (1+m_k)}{f_k \cdot N - 2r}\right| \nonumber \\
        &\le \frac{|\theta_A|(1-m_k) + |\theta_B|(1+m_k)}{f_k \cdot N - 2r} \nonumber\\
        &\le  \frac{1}{f_k \cdot N / (2r) - 1}.\nonumber
    \end{align} 
  Hence, we can guarantee that by at least $(1-\delta)^2$ probability,
    
    
    \begin{equation}
        \begin{cases}
          |f_k-f_k^*|_{\text{\textit{KVUE}}} \le \frac{e^\epsilon+2}{e^\epsilon-1}\sqrt{\frac{2}{N}\cdot \ln{\frac{2}{\delta}}},\\
          |m_k-m_k^*|_{\text{\textit{KVUE}}} \le \frac{(e^\epsilon+2)\cdot \sqrt{2\ln{ (2/\delta)}}}{(e^\epsilon -1)f_k \cdot \sqrt{N} - (e^\epsilon+2)\cdot \sqrt{2\ln{ (2/\delta)}}}, \nonumber
        \end{cases}
    \end{equation}
    
    
    and that by at least $(1-\delta)^2$ probability,
    
    \begin{equation}
        \begin{cases}
          |f_k-f_k^*|_{\text{\textit{KVOH}}} \le \frac{e^{\epsilon/2} + 1}{e^{\epsilon/2} - 1}\sqrt{\frac{2}{N}\cdot \ln{\frac{2}{\delta}}},\\
          |m_k-m_k^*|_{\text{\textit{KVOH}}} \le \frac{(e^{\epsilon/2}+1)\cdot \sqrt{2\ln{(2/\delta)}}}{f_k(e^{\epsilon/2}-1)\cdot \sqrt{N}-(e^{\epsilon/2}+1)\cdot \sqrt{2\ln{(2/\delta)}}}.\nonumber
        \end{cases}
    \end{equation}
    

\section{Conditional frequency and mean Estimation}
\label{sec: Conditional frequency and means}
    
    In this section, we present a complete analysis for privacy-preserving key-value data, which allows conditional analysis. Before giving the solutions, we first formulate the $L$-way conditional frequency and mean. To better understand the $L$-way conditional problems, we start from an example. For simplicity, we take $d = 3$ and $k \in \{\text{Hamburger, Fries, Pepsi}\}$ as a subset of Table~\ref{tbl: consumer kv data}. After discretization, each user's key-value data is listed as follows:
    
    \begin{table}[!htbp]
    \centering 
    \caption{Conditional table example} 
    \label{tbl: conditional table}
        \begin{tabular}{|p{1.5cm}<{\centering}|p{2cm}<{\centering}|p{1.5cm}<{\centering}|p{1.5cm}<{\centering}|}  
        \hline  
            \textbf{User} & $\langle k_{\text{Hamburger}}, v \rangle$ & $\langle k_{\text{Fries}}, v \rangle$ & $\langle k_{\text{Pepsi}}, v_c \rangle$  \\
            \hline
            \textbf{User1} & $\langle 1, 1 \rangle$ & $\langle 0, 0 \rangle$ & $\langle 1, -1 \rangle$  \\
            \textbf{User2} & $\langle 1,-1 \rangle$ & $\langle 1,1 \rangle$ & $\langle 1,1 \rangle$  \\
            \textbf{User3} & $\langle 0,0 \rangle$ & $\langle 1,-1 \rangle$ & $\langle 1,-1 \rangle$  \\
            \hline
        \end{tabular}
    \end{table}
    
    \begin{definition}[$L$-way Conditional Frequency and Mean]
    Given target key $k$ and $L$ conditional keys, the conditional frequency and conditional mean of key $k$ is defined as $f_{k|ck_1 = c_1, ..., ck_{L-1}=c_{L-1}}$ and $m_{k|ck_{1}=c_1, ..., ck_{L-1}=c_{L-1}}$, where $ck_i \in k_{[d]}$ represents a key and $c_i \in \{0,1\}$ represents the key $ck_i$ exists or not.
    \end{definition}
    
    Given conditions $\mathcal{C}: ck_1, ck_2, ..., ck_{L-1}= c_1, c_2, ..., c_{L-1}$, we say that a user meets conditions if the existence of key $ck_i$ is $c_i$. For example, $k_{\text{Fries}},k_{\text{Pepsi}}=0,1$ represents a consumer ordered Pepsi but not Fries (which is user1 in this example). With those $L-1$ conditions, we now formulate the $L$-way conditional frequency and means:
    
    \begin{equation}
        f_{k|ck_1, ..., ck_{L-1} = c_1, ..., c_{L-1}} = \frac{\vert \{u_i | u_i \in \mathcal{U}^{\mathcal{C}},\exists \langle k, v \rangle \in S_i\} \rvert }{\lvert \mathcal{U}^{\mathcal{C}} \rvert}, \nonumber
    \end{equation}
    
    \begin{equation}
        m_{k|ck_1, ..., ck_{L-1} = c_1, ..., c_{L-1}} = \frac{\sum_{u_i:u_i\in\mathcal{U}^{\mathcal{C}}, \langle k,v \rangle \in S_i} v}{\vert \{u_i | u_i \in \mathcal{U}^{\mathcal{C}}, \exists \langle k, v \rangle \in S_i\} \rvert } .\nonumber
    \end{equation}
    
    Where $\mathcal{U}^{\mathcal{C}}$ means users with conditions $\mathcal{C}$. For example, to represent consumer's average scores of Hamburgers among those who orders Pepsi, we can use the 2-way conditional mean $m_{k_\text{Hamburger}=1|k_{\text{Pepsi}}=1}$. The $L$-way conditional notions are easy to understand but not manageable. Considering that keys in conditions might be out-of-order, we introduce the $(\alpha, \beta)$-condition to formalize the $L$-way conditions to a length-$d$ bit vector:
    
    \begin{definition}[ $\alpha, \beta$-condition]
        Given conditions $\mathcal{C} = \{ck_1, ck_2, ck_L=c_1,c_2, ...,c_L\}$, $\alpha$ is used to represent what key is in condition, which is $\alpha_{i|k_i = ck_j}= 1$. And $\beta$ is used to represent the value of key, which is $\beta_i = c_j$ if there exists $j$ s.t. $ck_j = k_i$.
    \end{definition}
    
    For example, the conditions $\mathcal{C}= \{k_{\text{Hamburger}}=1, k_{\text{Fries}}=0\}$ can be represented by $\mathcal{C} = (\alpha=\textbf{11}0, \beta=\textbf{10}0)$. $\alpha=\textbf{11}0$ indicates that the first key and the second key is assigned in conditions, the conditional value of the key is in $\beta$, which is $\beta=\textbf{10}0)$. In the rest of this paper, we use $\mathcal{C} = (\alpha, \beta)$ for conditions representation.
    
    To handle the conditional estimation with privacy concerns, we first need to encode all of the key-value data. The proposed methods for frequency and mean estimation only works on one single key-value pair. To achieve $\epsilon$-LDP on the whole key-value pairs, each key-value pair should be encoded with $\epsilon' = \epsilon / d$, which might cause errors in the estimation results. To overcome this, we introduce the \textbf{I}ndexing \textbf{O}ne \textbf{H}ot encoding (\textit{IOH}) mechanism. Following \textit{KVOH}, a key-value pair $\langle k_i, v_i \rangle$ is first encoded to a single state by:
    \begin{equation}
        I(\langle k_i, v_i \rangle) = k_i \cdot \text{Discretization}(v_i) + 1.\nonumber
    \end{equation}
    
    Then we can get the index by all the $I(\langle k_i, v_i \rangle)$: 
    \begin{equation}
        I = \sum_{i=1}^d 3^{d-i} \cdot I(\langle k_i, v_i \rangle).\nonumber
    \end{equation}
    
    We initialize a zero array $\mathcal{A}$ and set $\mathcal{A}[I] =1$. The bit array is the one hot encoding of key-value pairs, like \textit{KVOH}, we can achieve $\epsilon$-LDP by using $\epsilon' = \epsilon/2$ on each bit. To sum up, the process is in Algorithm~\ref{alg: indexing one hot}. Unlike $KVOH$, the $IOH$ encoding mechanism encodes all of key-value of a user. For user1 in our example, the key-value pairs are first indexed with the indexing function:
    
    \begin{equation}
        \langle 1, 1 \rangle,\langle 0,0 \rangle,\langle 1,-1 \rangle \rightarrow 210_{(3)}.\nonumber
    \end{equation}
    
    \begin{algorithm}[t] 
    \caption{$IOH$: Indexing One Hot encoding} 
    \label{alg: indexing one hot}
    \begin{algorithmic}[1] 
        \REQUIRE ~~\\ 
            A user $u$'s set of key-value pairs $S = \{\langle k_1, v_1 \rangle, ..., \langle k_d, v_d \rangle \}$ (here $\langle k_j, v_j \rangle$ is set to $\langle 0, 0 \rangle$ if user $u$ does not have it); Privacy budget $\epsilon$.
        \ENSURE ~~\\ 
            $IOH(S, \epsilon$) is the perturbed key-value pair.
        \STATE Discretize each key-value data to $k_i', v_i'$;
        \STATE Indexing each encoded key-value pair by $k_i'\cdot v_i' + 1$ and get the overall index by:
            \begin{equation}
                I = (k_1'\cdot v_1' + 1) | (k_1'\cdot v_1' + 1) | ... | (k_d'\cdot v_d' + 1);
            \end{equation}
            
        \STATE Initialize a array: $\mathcal{A} = [0,0,...,0]$, where $|A| = 3^d$;
        \STATE Indexing: $\mathcal{A}[I] = 1$;
        \STATE Perturb each bit in $\mathcal{A}$ with probability $\epsilon/2$;
        \RETURN $\mathcal{A}$; 
    \end{algorithmic}
    \end{algorithm}

    The subscript $(3)$ here means the base of $210$ is 3. Thus user1 encodes his data to a bit array with its index 12 set to 1. When all the data are transferred to the aggregator, all the received bit vectors are adjusted and summed up to a $\mathcal{A}_s$:
    \begin{equation}
        \mathcal{A}_s[i] = \frac{(e^{\epsilon/2}+1)\cdot \sum_{j} \mathcal{A}_{j}[i] - N}{e^{\epsilon/2}-1}.
    \end{equation}
    
    Here, $\mathcal{A}_j$ denotes the $j$-th user's indexing one hot encoding vector. We will further extract the conditional frequency and conditional means from the summed array $\mathcal{A}_s[i]$. We will further use the adjusted $\mathcal{A}_{s}$ for conditional frequency estimation and conditional mean estimation.
    

\subsection{Conditional Frequency Estimation}
    
    To retrieve information from the summed array $A_s$, we first define the frequency counting operator.
    
    \begin{definition}[Frequency Counting Operator]
    \label{def: frequency counting operator}
        Given vector $\mathcal{A}_s$, the frequency counting operator $\mathcal{F}^\alpha_\beta[\mathcal{A}_s]: \mathbb{R}^{3^d} \rightarrow \mathbb{R} (\alpha,\beta \in \{0,1\}^d)$ counts the number of users with conditions $\alpha, \beta$.
    \end{definition}
    
    For example, if we want to know the number of users with $k_a, k_c = 1, 0$ in Table~\ref{tbl: conditional table}. We first get the $(\alpha, \beta) = (\textbf{1}0\textbf{1}, \textbf{1}0\textbf{0})$. We want to know $f_{k_a|k_c =1}$, thus we need to know the number of users with $k_a,k_c=1,1$ and the number of users with $k_c=1$. 
    
    \begin{align}
        \mathcal{F}^{\textbf{1}0\textbf{1}}_{\textbf{1}0\textbf{1}}[\mathcal{A}_s]= \mathcal{F}^{\textbf{111}}_{\textbf{101}} + \mathcal{F}^{\textbf{111}}_{\textbf{111}}.\nonumber
    \end{align}
    
    Another example is that $f_{k_c =0}$ can be represented by:
    \begin{align}
    \mathcal{F}^{00\textbf{1}}_{00\textbf{0}} & = \mathcal{F}^{0\textbf{11}}_{0\textbf{00}} + \mathcal{F}^{0\textbf{11}}_{0\textbf{10}} \nonumber\\
    & = \mathcal{F}^{\textbf{111}}_{\textbf{000}} + \mathcal{F}^{\textbf{111}}_{\textbf{100}} + \mathcal{F}^{\textbf{111}}_{\textbf{010}} + \mathcal{F}^{\textbf{111}}_{\textbf{110}}.\nonumber
    \end{align}
    
    To make it simple, when the superscript is all 1, we ignore it ($\mathcal{F}_\gamma$ is short for $\mathcal{F}_\gamma^{11...11}$ in the following equation). With the defined $\mathcal{F}$, we can calculate $\mathcal{F}_\beta^\alpha$ by:
    
    \begin{equation}
    \label{equ: counting operation}
    \mathcal{F}^\alpha_\beta[\mathcal{A}_s] = \sum\nolimits_{\gamma: \gamma \land \alpha = \beta} \mathcal{F}_{\gamma}[\mathcal{A}_s].
    \end{equation}
    
    The frequency counting operator can be used for the conditional frequency estimation. For example, the conditional frequency $f_{k_a|k_c =1}$ can be represented by:
    \begin{equation}
    f_{k_1|k_3=1} = \frac{\mathcal{F}^{101}_{101}[\mathcal{A}_s]}{\mathcal{F}^{001}_{001}[\mathcal{A}_s]}.\nonumber
    \end{equation}
    
    When encoding, all of the user's data are mapped into a length $3^d$ bit vector. For frequency estimation, we need to extract given key-value data under condition $\mathcal{C}$. We now introduce the notion of \textit{condition to frequency index} for computing $\mathcal{F}$. 
    
    \begin{definition}[Condition to Frequency Index]
        For conditional vector $\gamma \in \{0,1\}^d$, the corresponding index of $\mathcal{F}$ is given by $\mathcal{I}(\gamma) = \mathcal{I}(\gamma_0) | \mathcal{I}(\gamma_1) | ... | \mathcal{I}(\gamma_{d-1})$, where $\mathcal{I}(\gamma_i)$ is defined as: 
    \end{definition}
    
    \begin{equation}
        \mathcal{I}(\gamma_i) = \begin{cases}
        \{0,2\}, \quad &\text{If } \gamma_i = 1,\nonumber\\
        \{1\}, \quad &\text{If } \gamma_i = 0.\\
        \end{cases}
    \end{equation}
    
    For example, to compute $\mathcal{F}_{101}$, the index set is:
    \begin{align}
        \mathcal{I}(101) & = \mathcal{I}(1) | \mathcal{I}(0) | \mathcal{I}(1) \nonumber\\
        &= \{010, 012, 210, 212\}.\nonumber
    \end{align}
    
    With the index of $\mathcal{I}(\gamma)$, we can compute $\mathcal{F}_\gamma$ by:
    
    \begin{equation}
        \mathcal{F}_\gamma[\mathcal{A}_s] = \sum\nolimits_{i \in \mathcal{I}(\gamma)} \mathcal{A}_s[i]. 
    \end{equation}
    
    Following this example, the frequency under condition $\mathcal{C} = (\alpha, \beta)$ can be given by:
    
    \begin{equation}
        f_{k|\mathcal{C=(\alpha, \beta)}} =  \mathcal{F}_{\beta \lor \beta[k]=1}^{\alpha \lor \alpha[k]=1} / \mathcal{F}_\beta^\alpha.
    \end{equation}

\subsection{Conditional Mean Estimation}
    Like the frequency counting operator, we define two counting operations to handle the conditional mean estimation tasks. We use ${\mathcal{S}_k}|^\alpha_\beta$ to represent the sum of value with key $k$ under condition $\alpha, \beta$, Then the conditional mean can be given by:
    \begin{equation}
    \label{equ: condition mean main}
        m_{k|\mathcal{C} =(\alpha, \beta)} = \frac{\mathcal{S}_{k|\mathcal{C} \lor \{{k=1}\}}}{\mathcal{F}^{\alpha \lor \alpha[k]=1}_{\beta \lor \beta[k]=1}}.
    \end{equation}
    
    The notion $\alpha \lor \alpha[k] =1$ and $\alpha \lor \alpha[k]=1$ indicates that when considering $m_{k|\mathcal{C}=(\alpha, \beta)}$, the key $k$ should be included. The main problem now is to calculate $\mathcal{S}_{k|\mathcal{C}}$. Like $\mathcal{F}^{\alpha}_{\beta}$, we use $\mathcal{S}_{k,\gamma}$ to represent $(\mathcal{S}_k)^{11...11}_\gamma$. For the $k$-th key-value pair, the value can be $1$ and $-1$. The mean estimation then turns to be the counting problem: to count the number of key-value pairs with the $k$-th value be 1 and be -1. The symbol $\mathcal{S}_{k,\gamma}$ calculates the sum of values with key $k$. After discretization, the sum of $\mathcal{S}_{k,\gamma}[\mathcal{A}_s]$ can be divided into two parts: those with the $k-$th key-value pair being $\langle 1, 1 \rangle$ (denoted as $\mathcal{S}_{k, \gamma}^+[\mathcal{A}_s]$) and those being $\langle 1, -1\rangle$ (denoted as $\mathcal{S}_{k, \gamma}^-[\mathcal{A}_s]$). 
    
    \begin{align}
        \mathcal{S}_{k|\mathcal{C}} &= \mathcal{S}_{k|\mathcal{C}}^+[\mathcal{A}_s] - \mathcal{S}_{k|\mathcal{C}}^-[\mathcal{A}_s] \nonumber\\
        &= \sum\nolimits_{\gamma: \gamma \land \alpha = \beta} \mathcal{S}_{k,\gamma}^+[\mathcal{A}_s] - \sum\nolimits_{\gamma: \gamma \land \alpha = \beta} \mathcal{S}_{k,\gamma}^-[\mathcal{A}_s].
    \end{align}
    
     Like the frequency index operator, we now define the mean index operator to extract $\mathcal{S}_{k,\gamma}^+$ and $\mathcal{S}_{k,\gamma}^-$ from $\mathcal{A}_s$.
    
    \begin{definition}[condition to mean index]
        For conditional vector $\gamma \in \{0,1\}^d$, the corresponding index of $\mathcal{S}_{k,\gamma}^+[\mathcal{A}_s]$ (and $\mathcal{S}_{k,\gamma}^-[\mathcal{A}_s]$) can be represented by: 
        
        \begin{equation}
            \mathcal{S}_{k,\gamma}^+[\mathcal{A}_s] = \mathcal{I}(\gamma_0) | ... | \mathcal{I}(\gamma_{k-1}) | \underbrace{\mathcal{I}^+(\gamma_{k})}_{\text{value 1}} | \mathcal{I}(\gamma_{k+1}) | ... | \mathcal{I}(\gamma_{d-1}), \nonumber
        \end{equation}
        \begin{equation}
            \mathcal{S}_{k,\gamma}^-[\mathcal{A}_s] = \mathcal{I}(\gamma_0) | ... | \mathcal{I}(\gamma_{k-1}) | \underbrace{\mathcal{I}^-(\gamma_{k})}_{\text{value -1}} | \mathcal{I}(\gamma_{k+1}) | ... | \mathcal{I}(\gamma_{d-1}), \nonumber
        \end{equation}
  where the $\mathcal{I}^+(\gamma)$ and $\mathcal{I}^-(\gamma)$ are defined as:
        
    \end{definition}
    
    \begin{equation}
            \mathcal{I}^+ (\gamma) = \begin{cases}
            \{2\}, \quad &\text{If } \gamma = 1,\nonumber\\
            \{1\}, \quad &\text{If } \gamma = 0,\\
            \end{cases}
        \end{equation}
        
        \begin{equation}
            \mathcal{I}^- (\gamma) = \begin{cases}
            \{0\}, \quad &\text{If } \gamma = 1,\nonumber\\
            \{1\}, \quad &\text{If } \gamma = 0.\\
            \end{cases}
        \end{equation}
    
    The only difference between conditional frequency and conditional mean is that for key $k$, the frequency estimation needs the overall number of value $1$ and value $-1$. Whereas for mean estimation, we need to estimate those with value $1$ and those with value $-1$ separately.

    \begin{figure}[t]
    \centering
    \footnotesize
    \begin{tabular}{cc}
        \vspace{-8pt}
        \hspace{-6mm}\includegraphics[width=0.3\textwidth]{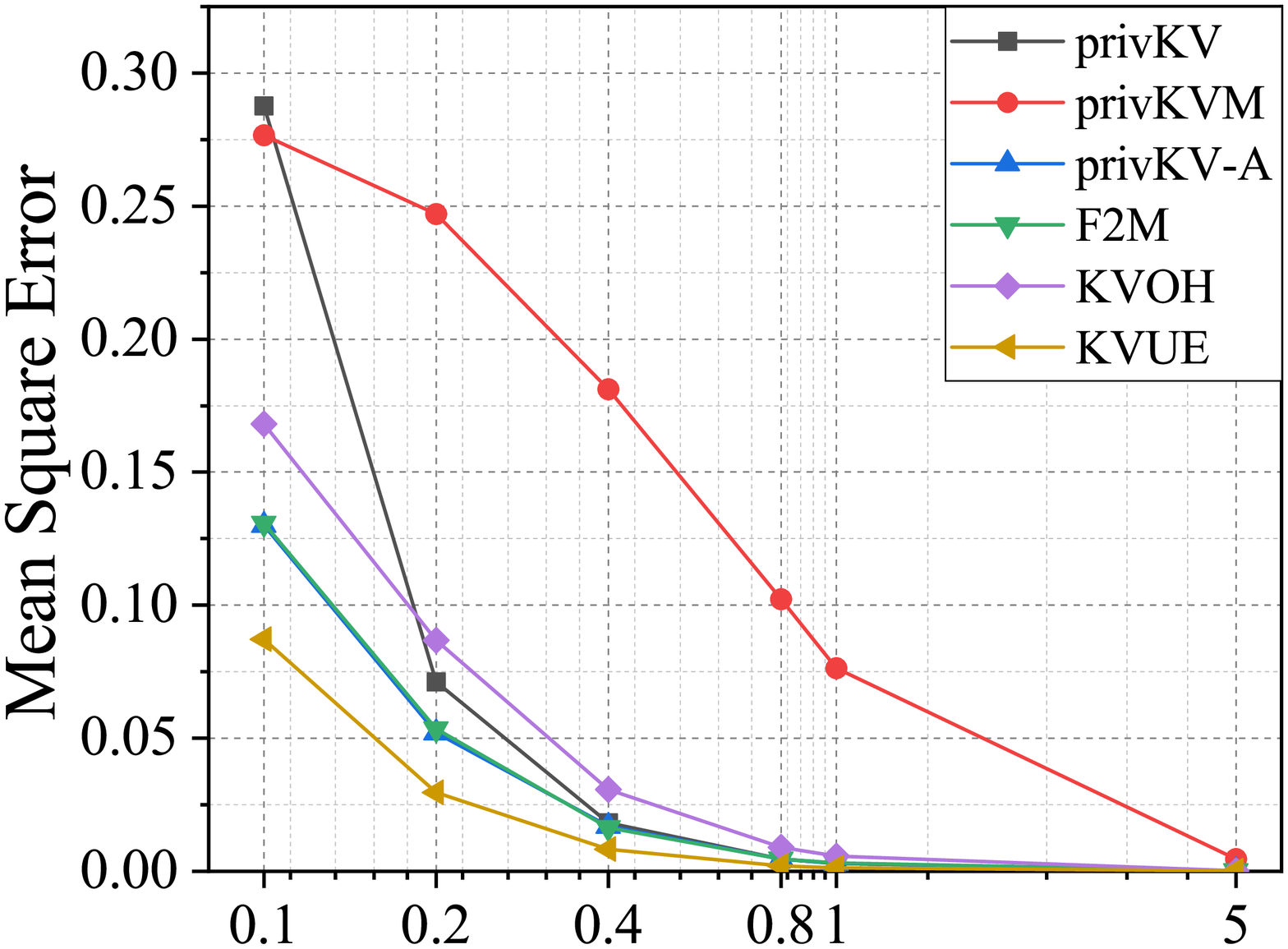} &
        \hspace{-12mm}\includegraphics[width=0.3\textwidth]{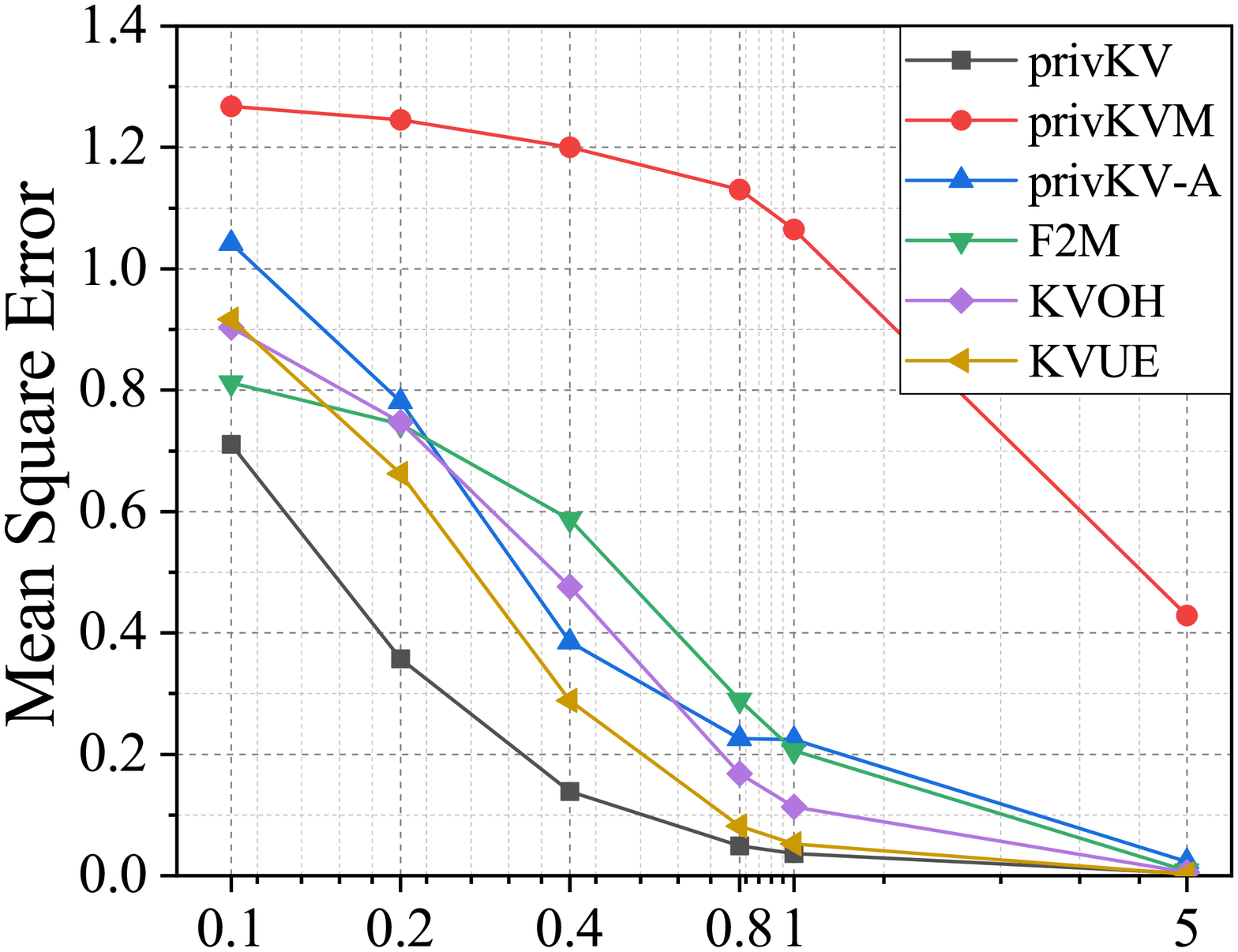}
        \\
        \hspace{-6mm} (a) MovieLens-frequency & 
        \hspace{-12mm} (b) MovieLens-mean 
        \vspace{-10pt}
        \\
        \vspace{-10pt}
        \hspace{-6mm}\includegraphics[width=0.3\textwidth]{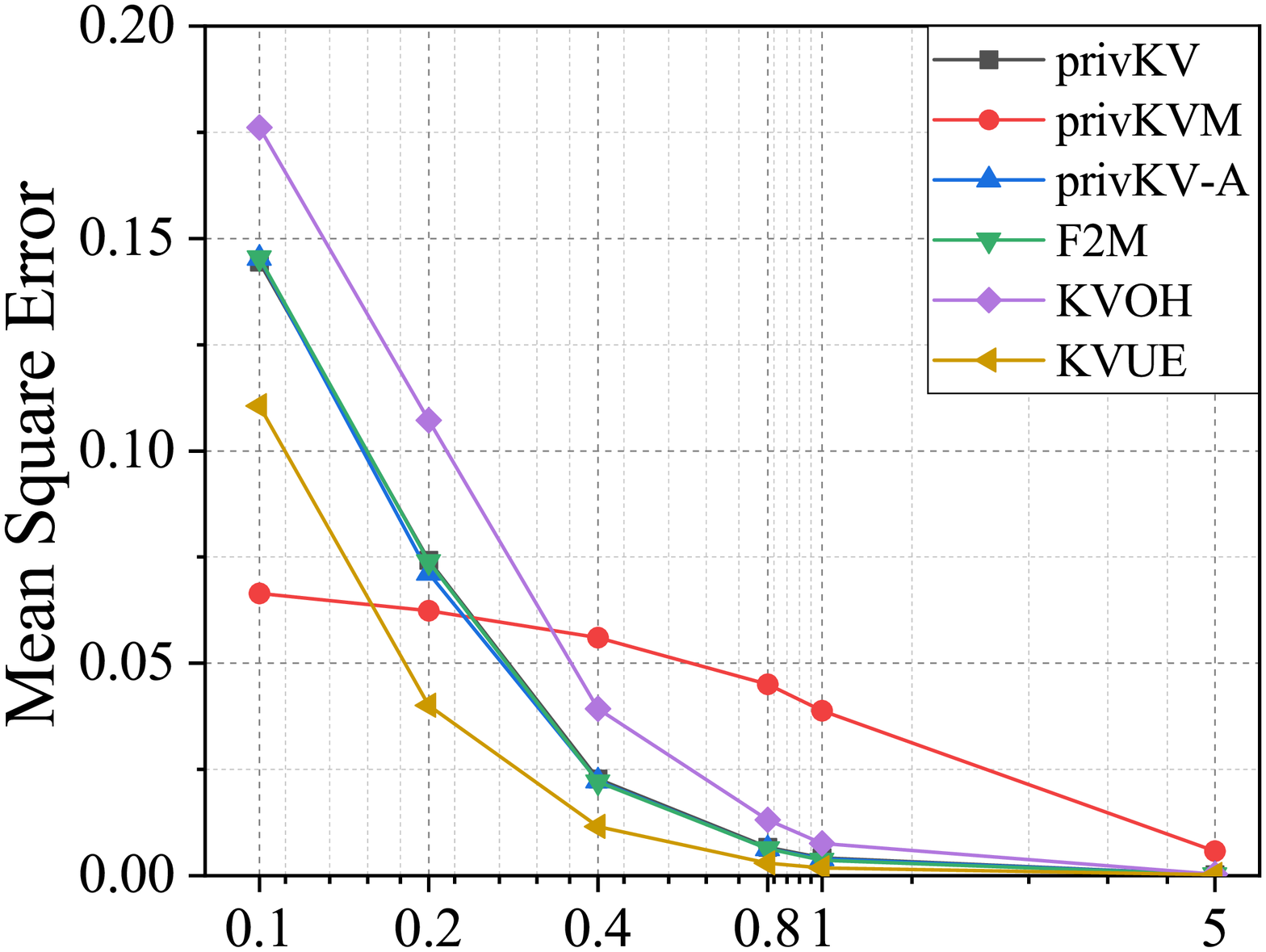} &
        \hspace{-12mm}\includegraphics[width=0.3\textwidth]{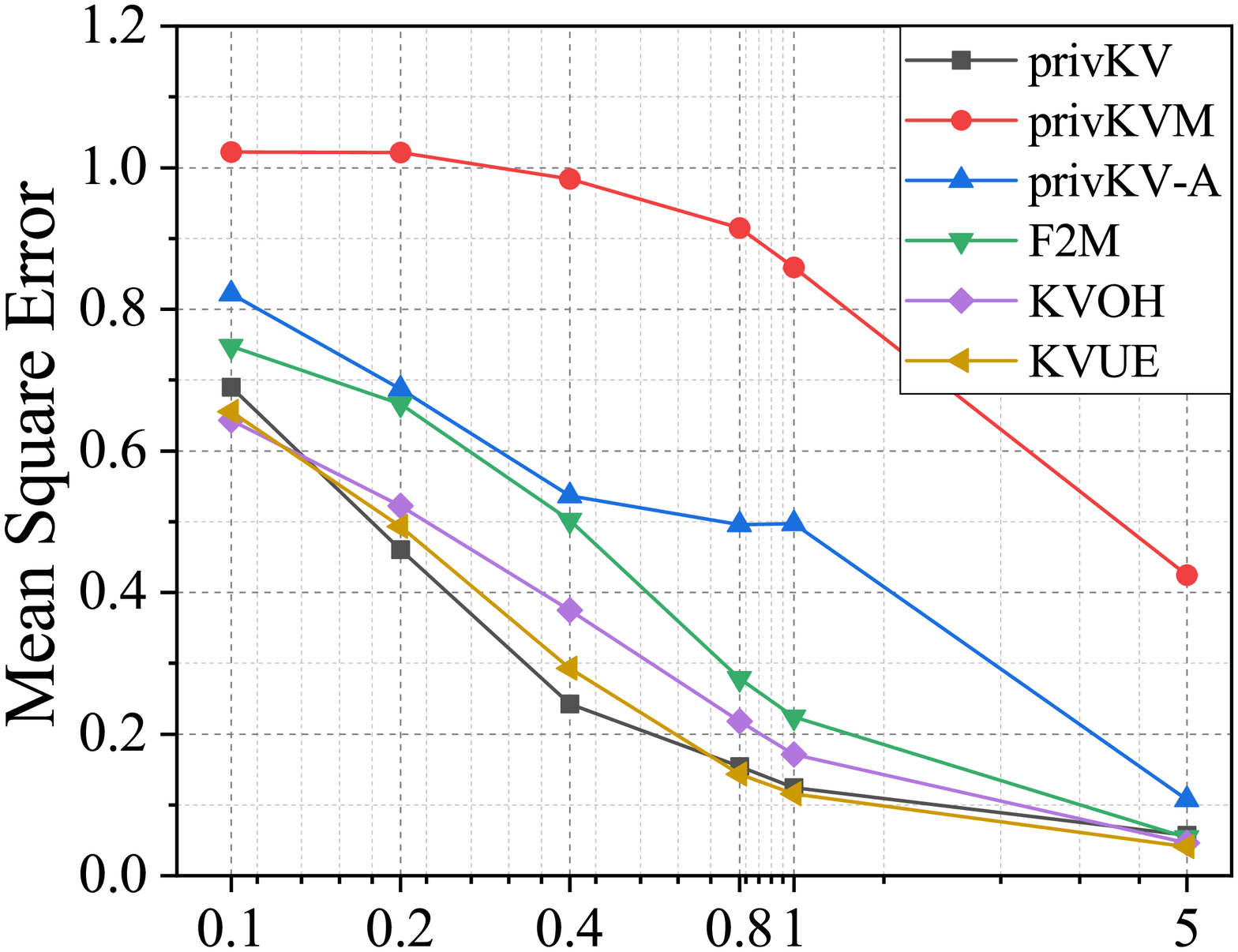}
        \\
        \hspace{-6mm} (c) Gaussian-frequency & 
        \hspace{-12mm} (d) Gaussian-mean 
        \vspace{-10pt}
        \\
        \vspace{-10pt}
        \hspace{-6mm}\includegraphics[width=0.3\textwidth]{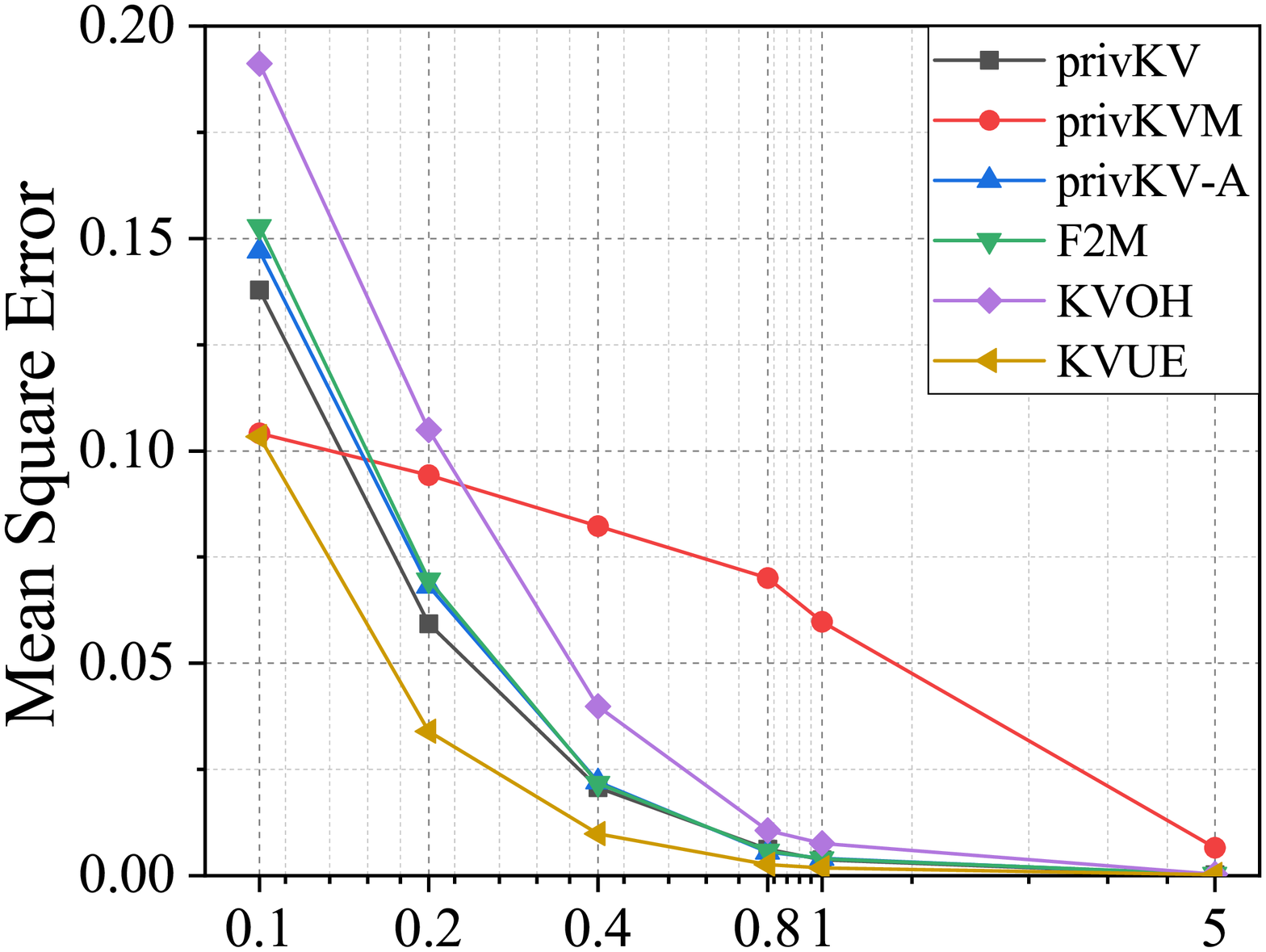} &
        \hspace{-12mm}\includegraphics[width=0.3\textwidth]{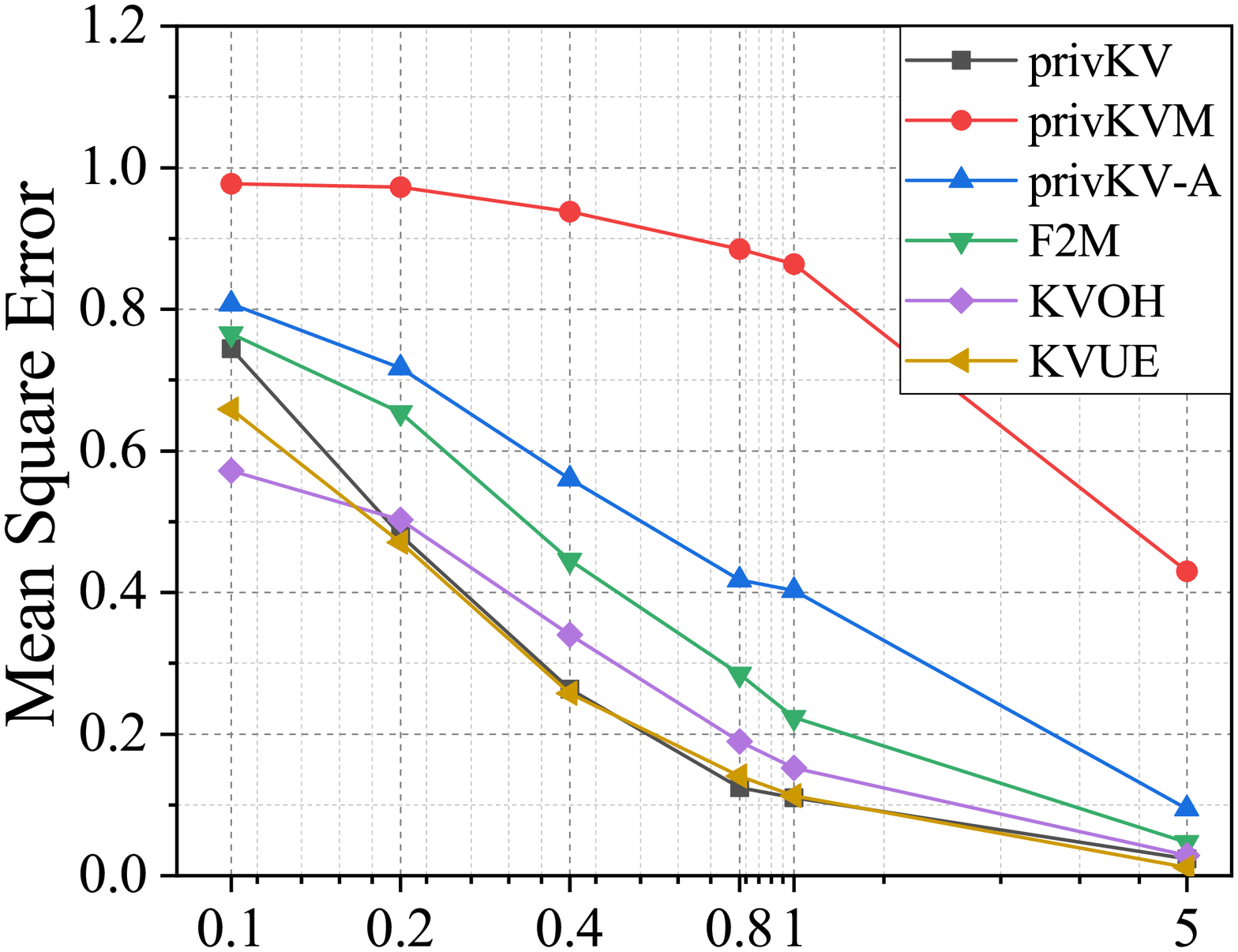}
        \\
        \hspace{-6mm} (e) Uniform-frequency & 
        \hspace{-12mm} (f) Uniform-mean 
    \end{tabular}
    \caption{Overall results of estimation varying $\epsilon$.}
    \label{fig: overall result}
    \end{figure}

\section{Analysis and Evaluation}
\label{sec: experiments}
    
    In this section, we empirically evaluate the performance of proposed mechanisms. The \textit{PrivKV}-based mechanisms~\cite{qingqing2019privkv} have shown great advantages in frequency estimation over proposed mechanisms like \textit{RAPPOR}~\cite{erlingsson2014rappor}, \textit{k-RR}~\cite{kairouz2014extremal} and \textit{SHist}~\cite{bassily2015local}. That is the same in mean estimation, over \textit{Harmony}~\cite{NguyenSS16collect} and \textit{MeanEst}~\cite{duchi2014privacy}. Thus we only  compare our proposed mechanisms with \textit{PrivKV}-based mechanisms, namely \textit{PrivKV} and \textit{PrivKVM}.
    
    \textbf{Datasets used.} We evaluate the proposed methods over a real-world dataset and synthetic datasets. We first use the \textbf{MovieLens} dataset~\cite{harper2016movielens}. This dataset samples were collected by the GroupLens Research Project. It contains over 20\textit{M} ratings from 138,000 users on over 27,000 movies. Each user has rated at least 20 movies. For each anonymous person, ratings are treated as key-value data. We first exact the top-100 most rated movies as our key space $\mathcal{K}$ and extract s smaller dataset. We also generated two synthetic datasets: the \textbf{Uniform} dataset and the \textbf{Gaussian} dataset. The frequency and mean for different keys follow the uniform distribution and Gaussian distribution. Each generated dataset has 100 keys and 100,000 records. 
    
    \textbf{Default parameters and settings.} In the frequency and mean estimation experiment, we acquire the distributions of estimation error by repeatedly encoding and decoding 50 times in each experimental instance. Each user randomly picks up one key-value pair and encodes with different mechanisms. Then the aggregator decodes with the corresponding mechanism. When encoding, the privacy budget varies from 0.1 to 5. For the \textit{PrivKVM}, we set the iterations to be 10. The result is measured with \textit{AE} (Absolute Error) and \textit{MSE} (Mean Square Error). For the \textit{F2M} estimator, we set the default $\overline{v}=1$. Also, the influence of the default value is discussed in Section VI-C.
    
    \begin{figure*}[t]
    \centering
    \footnotesize
    \begin{tabular}{cccc}
        \vspace{-10pt}
        \hspace{-8mm}\includegraphics[width=0.3\textwidth]{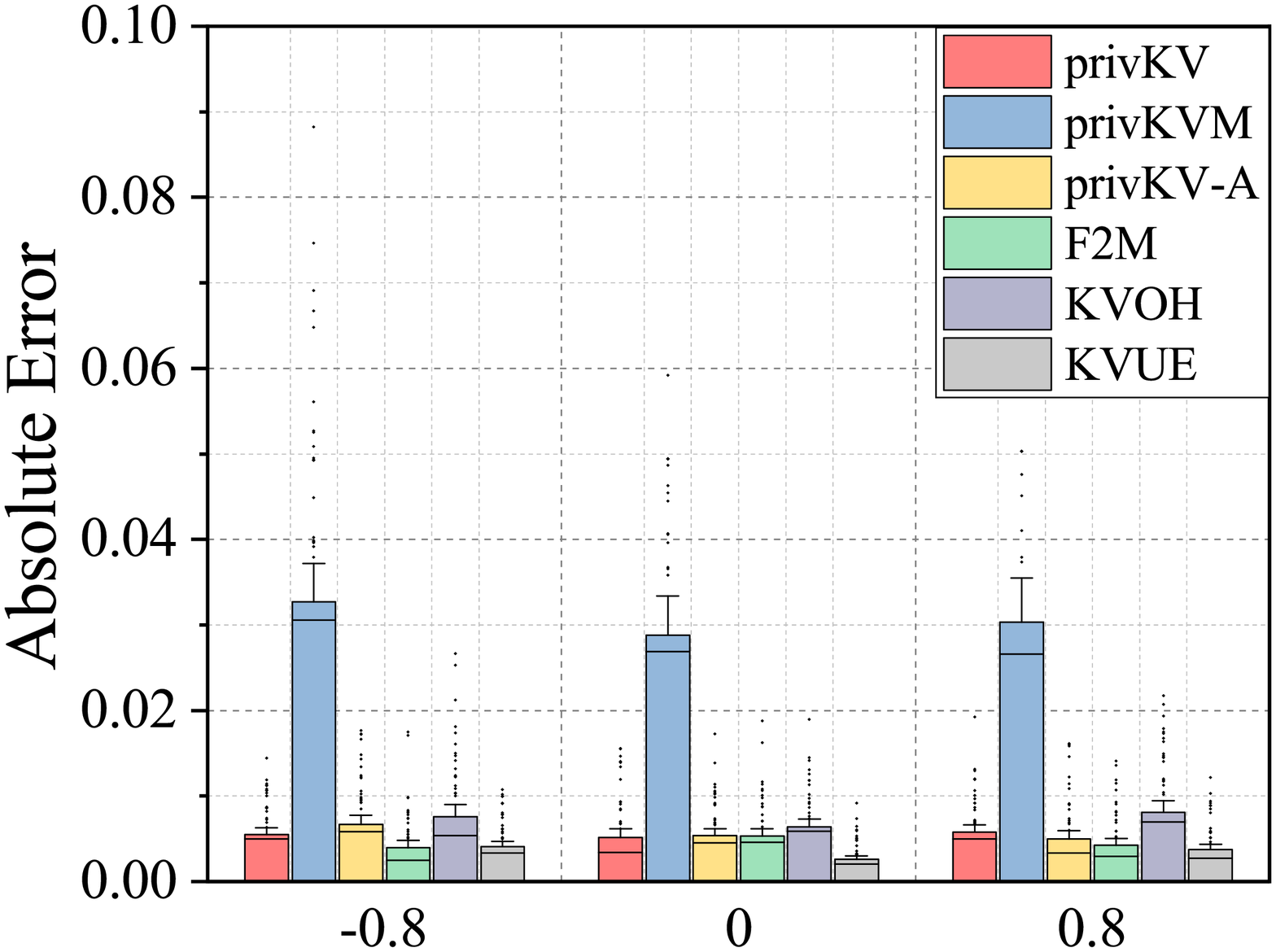} &
        \hspace{-12mm}\includegraphics[width=0.3\textwidth]{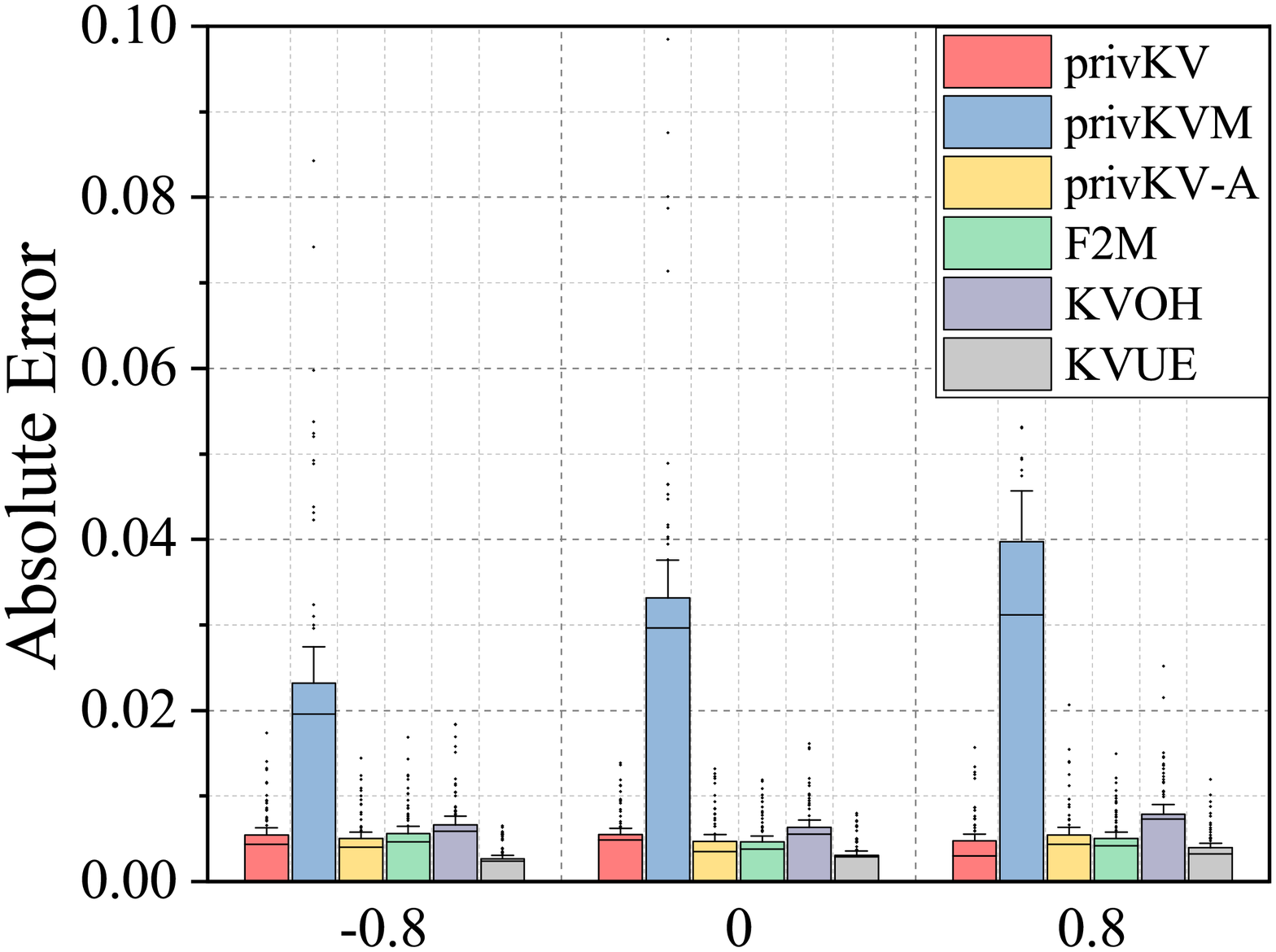} &
        \hspace{-12mm}\includegraphics[width=0.3\textwidth]{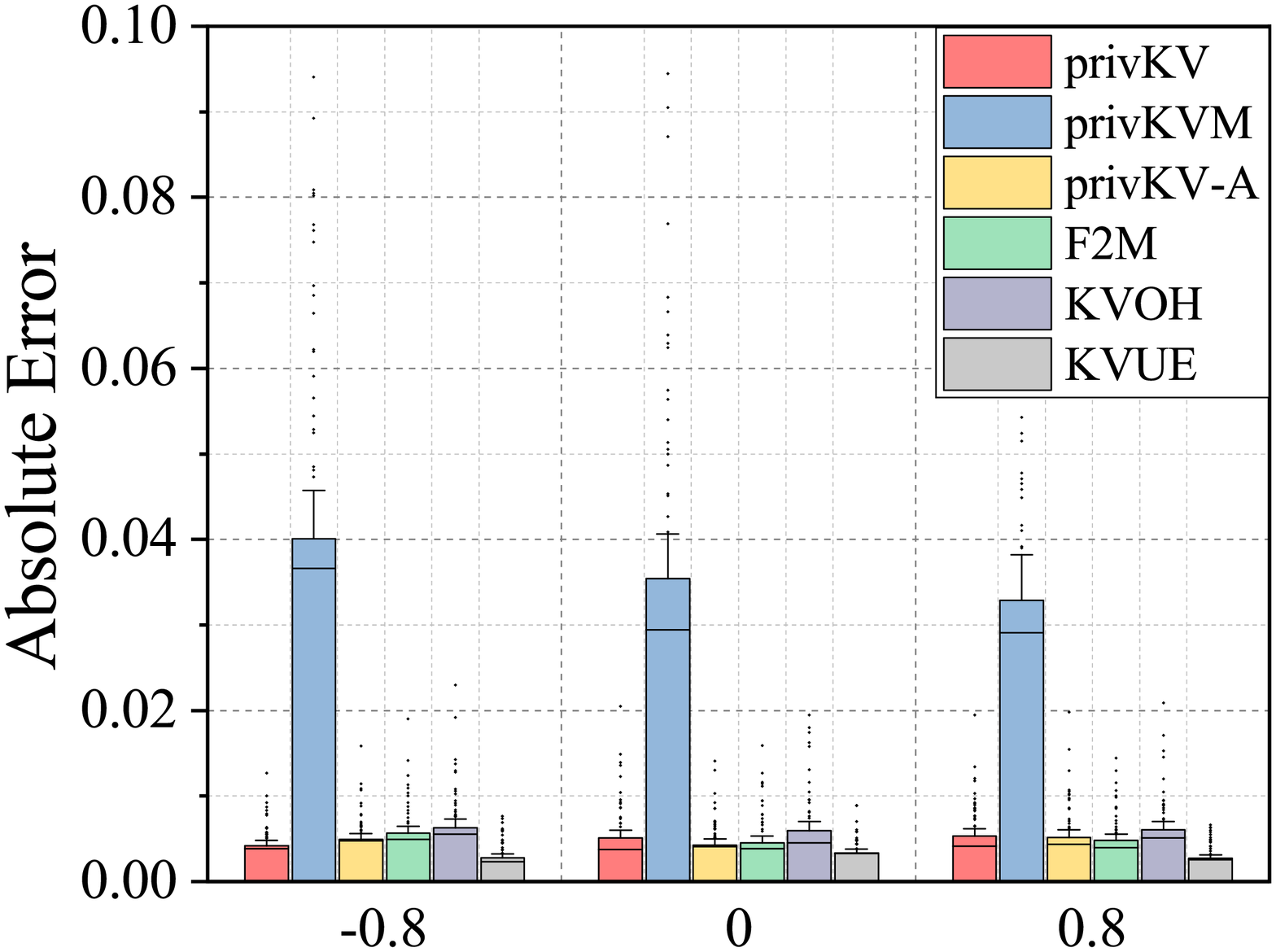}& 
        \hspace{-12mm}\includegraphics[width=0.3\textwidth]{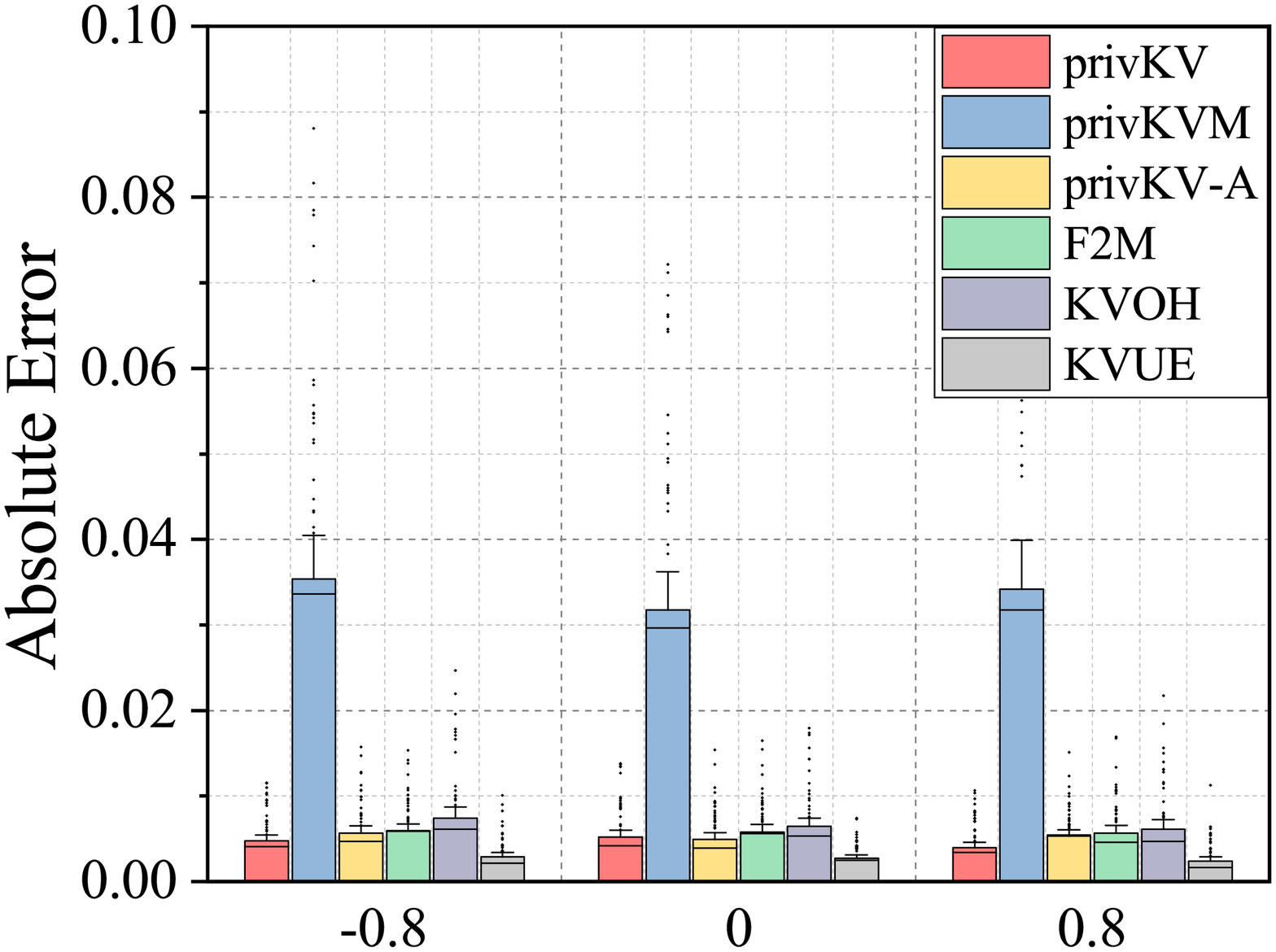} \\
        \hspace{-8mm} (a) extreme low frequency & 
        \hspace{-12mm} (b) low frequency & 
        \hspace{-12mm} (c) average frequency & 
        \hspace{-12mm} (d) high frequency 
    \end{tabular}
    \caption{Frequency estimation error of different mechanisms on Gaussian distribution data, with respect to the combinations of extreme low frequency, low frequency, average frequency, high frequency, and low average, middle average, high average.}
    \label{fig:gauss_specially_f_resout}
    \end{figure*}

    \begin{figure*}[t]
    \centering
    \footnotesize
    \begin{tabular}{cccc}
        \vspace{-10pt}
        \hspace{-8mm}\includegraphics[width=0.3\textwidth]{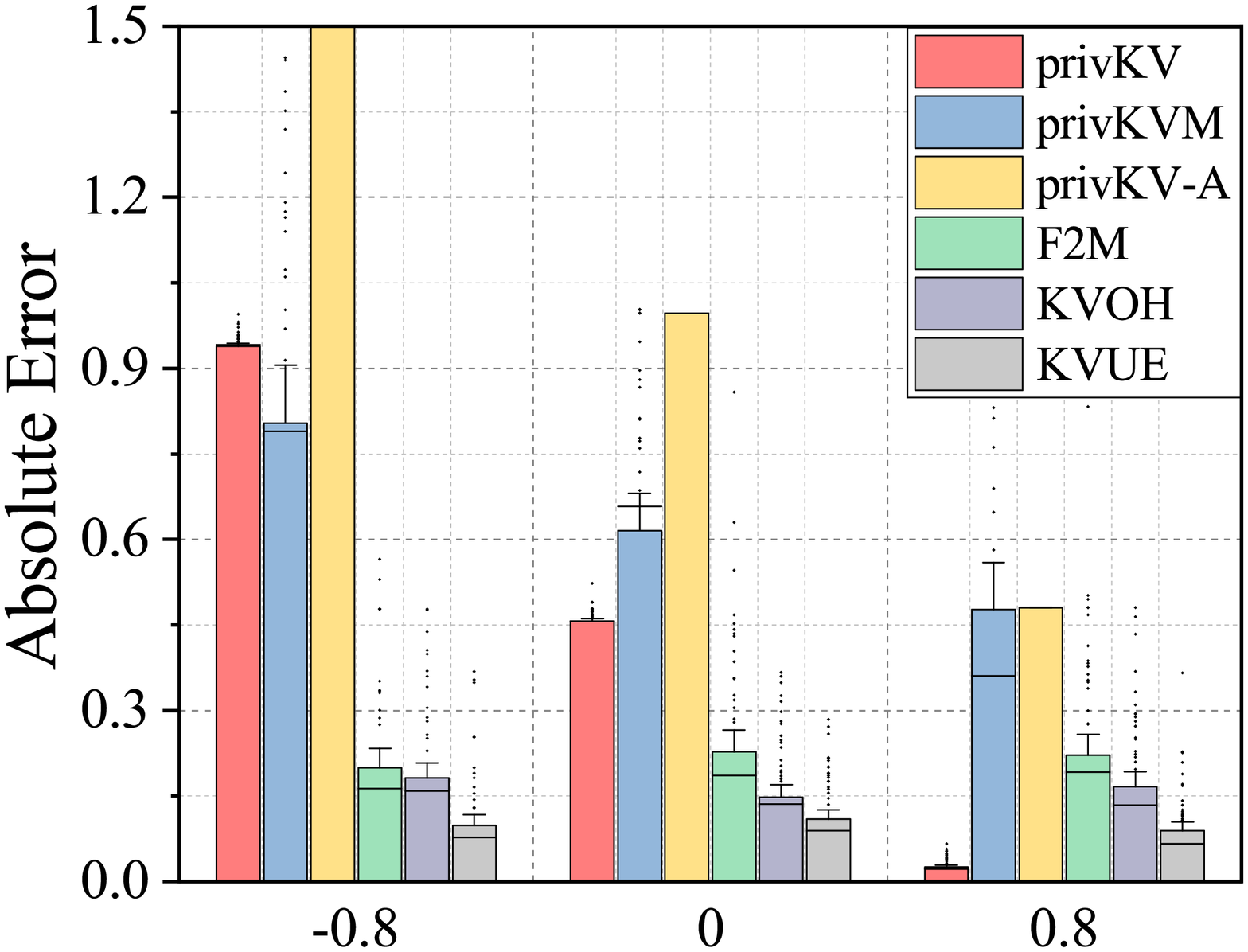} &
        \hspace{-12mm}\includegraphics[width=0.3\textwidth]{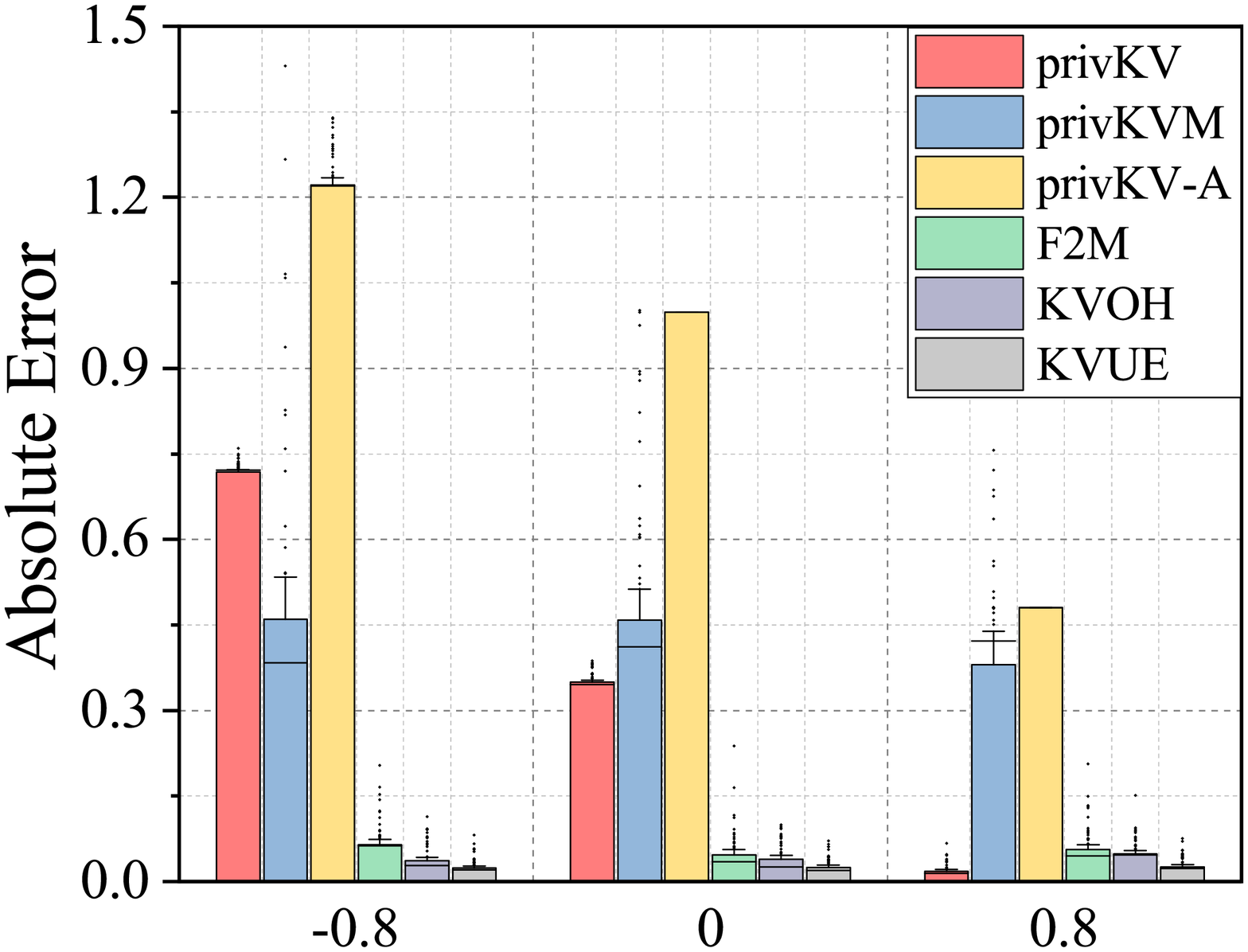} &
        \hspace{-12mm}\includegraphics[width=0.3\textwidth]{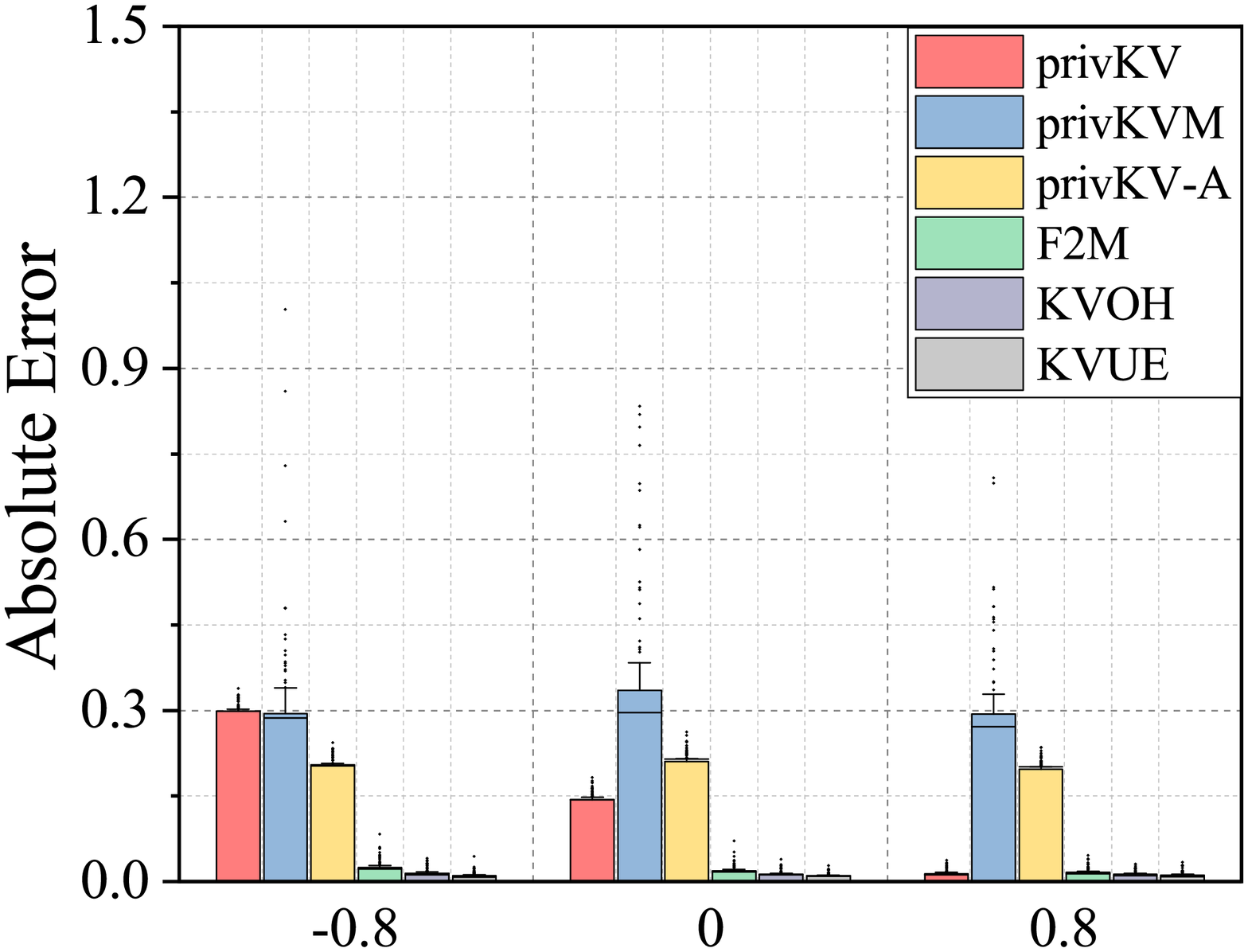} & 
        \hspace{-12mm}\includegraphics[width=0.3\textwidth]{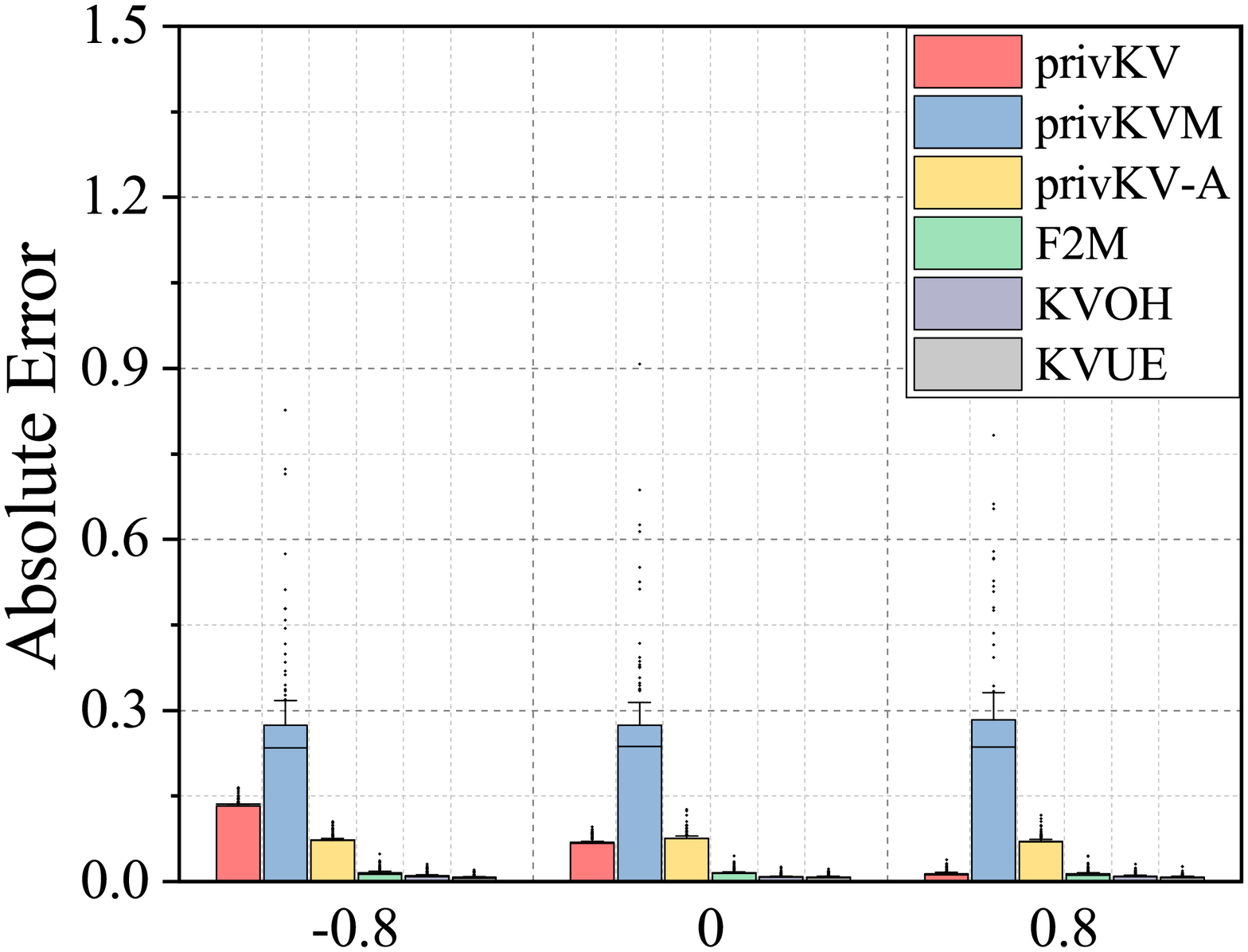} \\
        \hspace{-8mm} (a) extreme low frequency & 
        \hspace{-12mm} (b) low frequency & 
        \hspace{-12mm} (c) average frequency & 
        \hspace{-12mm} (d) high frequency 
    \end{tabular}
    \caption{Mean estimation error of different mechanisms on Gaussian distribution data, with respect to the combinations of extreme low frequency, low frequency, average frequency, high frequency, and low average, middle average, high average.}
    \label{fig:gauss_specially_m_resout}
    \end{figure*}

\subsection{Overall Results}
    
    We first list the theoretical communicating cost between a user end to the aggregator end. The cost is based on the number of state of the encoded key-value data. For example, the encoded space of \textit{PrivKV} is $\langle k', v' \rangle \in \{\langle 0, 0\rangle, \langle 1, 1\rangle, \langle 1, -1\rangle\}$, thus the communicating cost for key-value encoded by \textit{PrivKV} can be compressed to $\log_2 3$. The \textit{PrivKVM} works in an iterative way. Thus the communication cost is $c$ times that of \textit{PrivKV}, where $c$ is the number of iterations. When encoding, a user needs to pick up one key from the key space $\mathcal{K}$. Thus an index should also be sent to the aggregator. The cost for index is $\log_2 |\mathcal{K}|$. The costs of different mechanisms are listed in Table~\ref{tbl: cost}.
    
    \begin{table}[!htbp]
    \centering 
    \caption{Communication cost for one key-value data}
    \label{tbl: cost} 
    \begin{tabular}{|c|c|c|c|c|}  
        \hline 
        \textbf{Methods} & \textit{PrivKV} & \textit{F2M} & \textit{KVUE} & \textit{KVOH}\\
        \hline
        \textbf{Cost} & $\log_2 (3 |\mathcal{K}|)$ & $2\cdot \log_2 |\mathcal{K}|$ & $\log_2 (3 |\mathcal{K}|)$ & $3 \cdot \log_2|\mathcal{K}|$ \\
        \hline
    \end{tabular}
    \end{table}
    
    Figure~\ref{fig: overall result} plots the estimation errors of different mechanisms with different privacy budgets. Among all these six mechanisms, the \textit{PrivKVM} is the only one that outputs an unbiased mean estimation. However, our simulations indicate the effectiveness of both frequency and mean estimation. As the \textit{PrivKVM} achieves unbiased estimation by iterating with the aggregator, and in each round, the privacy budget is very small ($\epsilon' = \epsilon / c$). Thus estimation error in each round accumulates.  When the privacy budget is not very small ($\epsilon>0.4$), the \textit{KVOH}, \textit{KVUE}, \textit{KVOH}, \textit{PrivKV} and \textit{PrivKV-A} can achieve estimation error under $0.05$. Over the tested mechanisms,   $KVUE$   achieves lower estimation error considering different privacy-preserving levels on both generated dataset and real-world dataset.
    
    All of these mechanisms have higher mean estimation errors compared with frequency estimation. We think it is because of the natural insufficiency of local differential privacy: the estimation accuracy is influenced by the volume of data. When estimating the frequency, we need to estimate the number of key-value data with key from $N$ users, which is $N\cdot f_k$ from $N$. Compared with that, the mean estimation task requires estimating the number of key-value pair with value 1 and value $-1$ from the estimated key-value data with key. Thus the accuracy of key estimation affects the performance of mean estimation. Like frequency estimation, generally, \textit{KVUE} achieves lowest estimating error.
    
\subsection{Scalability}

    \begin{figure}[t]
    \vspace{-10pt}
    \centering
    \footnotesize
    \begin{tabular}{cc}\footnotesize
        \hspace{-7mm}\includegraphics[width=0.3\textwidth]{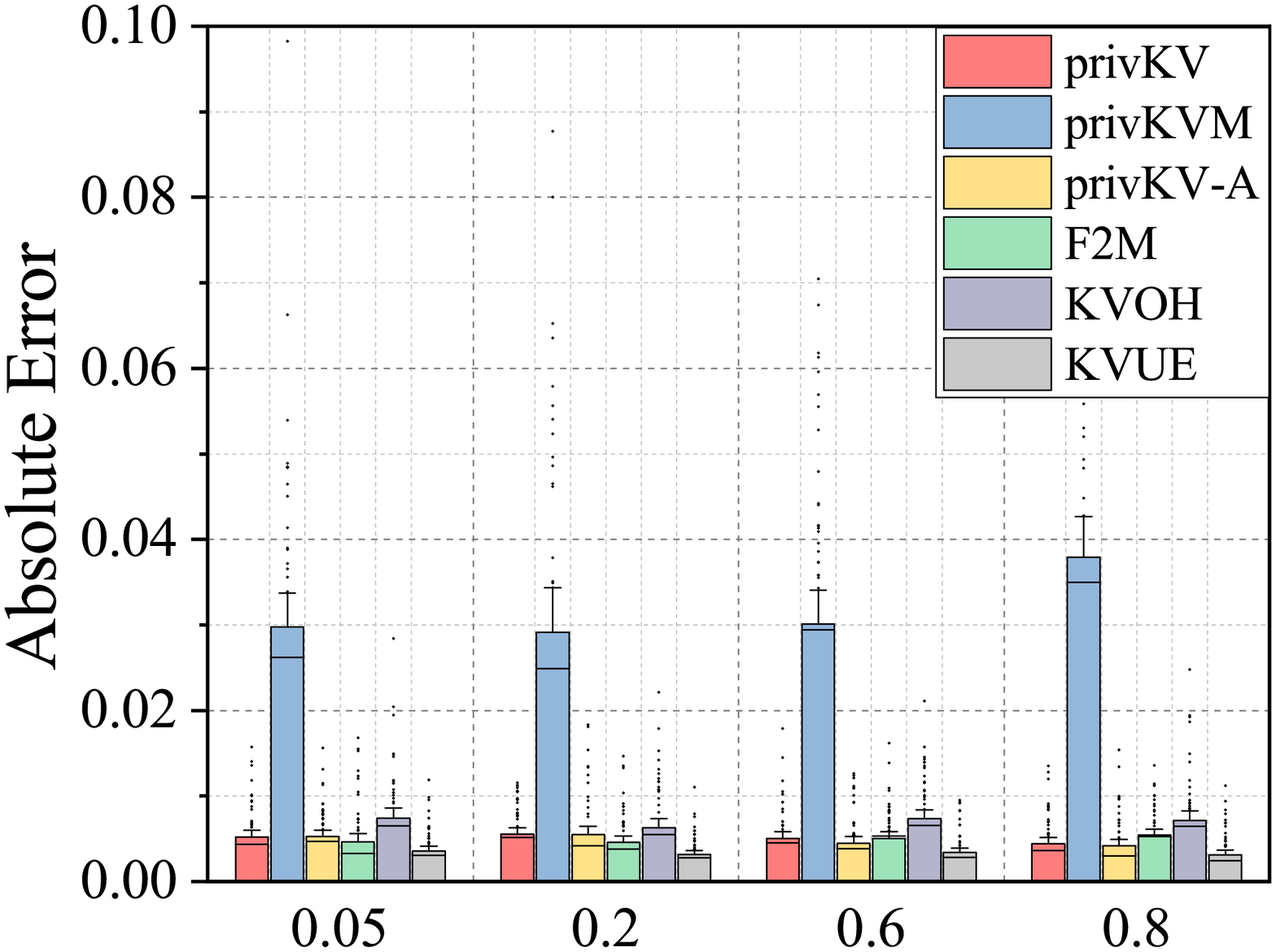} &
        \hspace{-12mm}\includegraphics[width=0.3\textwidth]{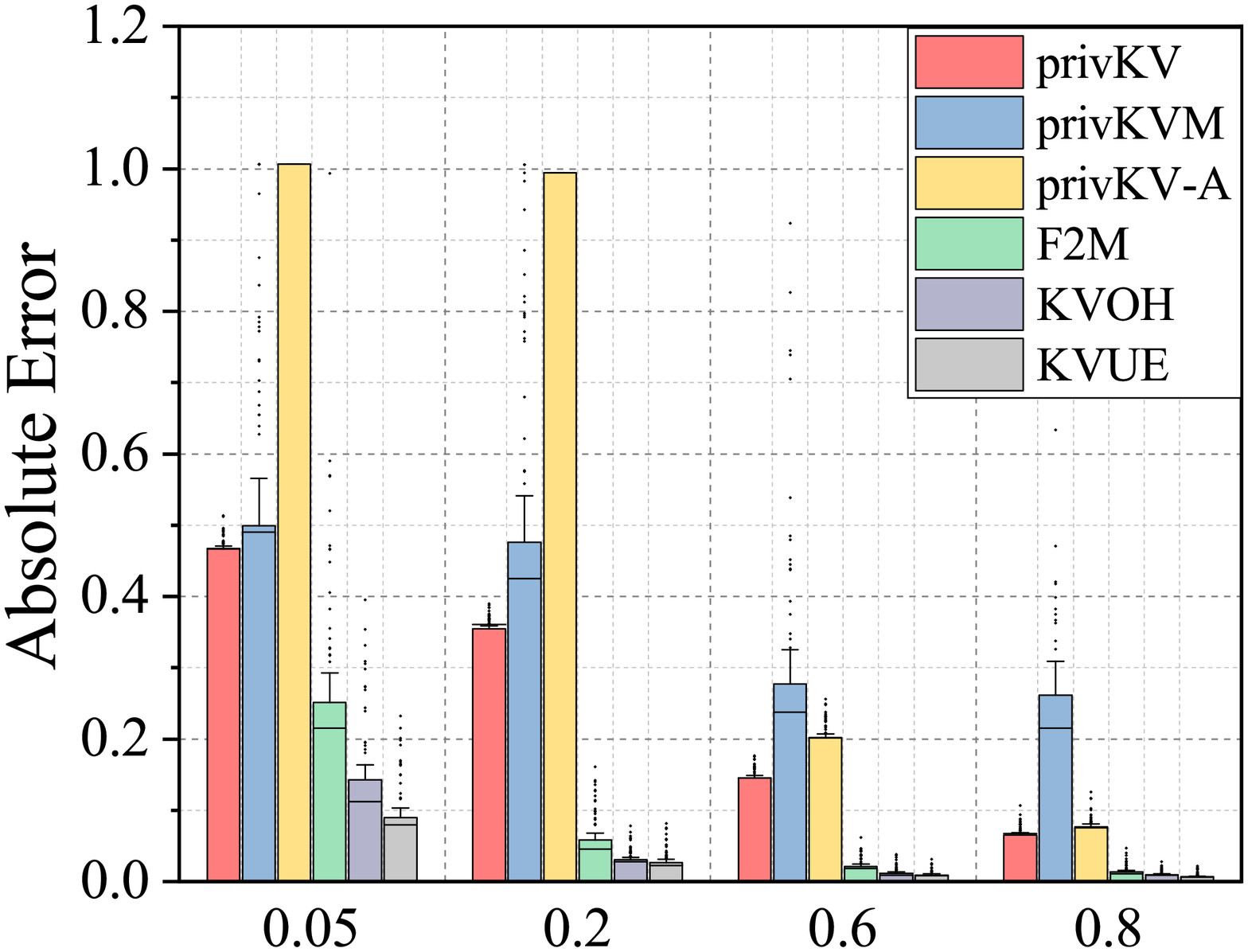}
        \\
        \hspace{-7mm} (a) frequency estimation & 
        \hspace{-12mm} (b) mean estimation 
    \end{tabular}
    \caption{Estimation error on uniform distribution.}
    \vspace{-20pt}
    \label{fig:uniform_specially_resout}
    \end{figure}
    
    In this section, we evaluate the performance of estimating mechanisms on different circumstances. For the frequency estimation, We divide the frequency into four situations: \textit{extreme low frequency} with $f_k = 0.05$, \textit{low frequency} with $f_k = 0.2$, \textit{middle frequency} with $f_k = 0.6$ and \textit{high frequency} $f_k = 0.8$. For the mean estimation, we divide the mean into three situations: \textit{low average} with $m_k$ around $-0.8$, \textit{middle average} with $m_k$ around $0$ and \textit{high average} with $m_k$ around $0.8$. We generate several these kinds of dataset with Gaussian distribution and uniform distribution. Each generated dataset contains 100,000 key-value pairs. 

    Figure~\ref{fig:gauss_specially_f_resout} and Figure~\ref{fig:gauss_specially_m_resout} show the box-plot of frequency estimation and mean estimation results with Gaussian distribution. It turns out that estimation errors of different mechanisms are not under the influence of frequencies. However, \textit{PrivKV}, \textit{PrivKVM} and \textit{PrivKV-A} are susceptible to the location of means. These three mechanisms achieve higher estimation accuracy with the rise of mean. In both cases, \textit{PrivKVM} returns an inaccurate result with large variance. As we analyzed formally, the error of estimation might accumulate when iterating. Compared with other mechanisms, the \textit{KVUE} mechanism achieves the lowest error in both frequency estimation and mean estimation. Also, \textit{F2M} and \textit{KVOH} mechanisms attain acceptable results compared with existing methods. Also, with the increase of frequency, the variance of error decreases. This is because, with more usable data, the estimation becomes settled.
    
    When estimating with uniform distributed data, the result is shown in Figure~\ref{fig:uniform_specially_resout}. As in the case of Gaussian distribution, the result of frequency is not profoundly affected by situations of frequency. Like aforementioned, we can draw that a higher frequency leads to a lower mean estimation error.

\subsection{Influence of default value in \textit{F2M}}

    In the \textit{F2M} mechanism, we set the default value of encoding to $\overline{v} = 1$. We think that by setting the default value to 1, the discretized value is always the same as 1. That avoids additional errors for further estimation. Thus, a natural question occurs. Will the value of $\overline{v}$ influence the performance of mean estimation? Here, we do not need to discuss the impact of the default value to frequency estimation as setting default only affects the process of mean estimation.
    
    \begin{figure}[t]
    \centering
    \footnotesize
    \begin{tabular}{cc}\footnotesize
        \vspace{-10pt}
        \hspace{-10mm}\includegraphics[width=0.3\textwidth]{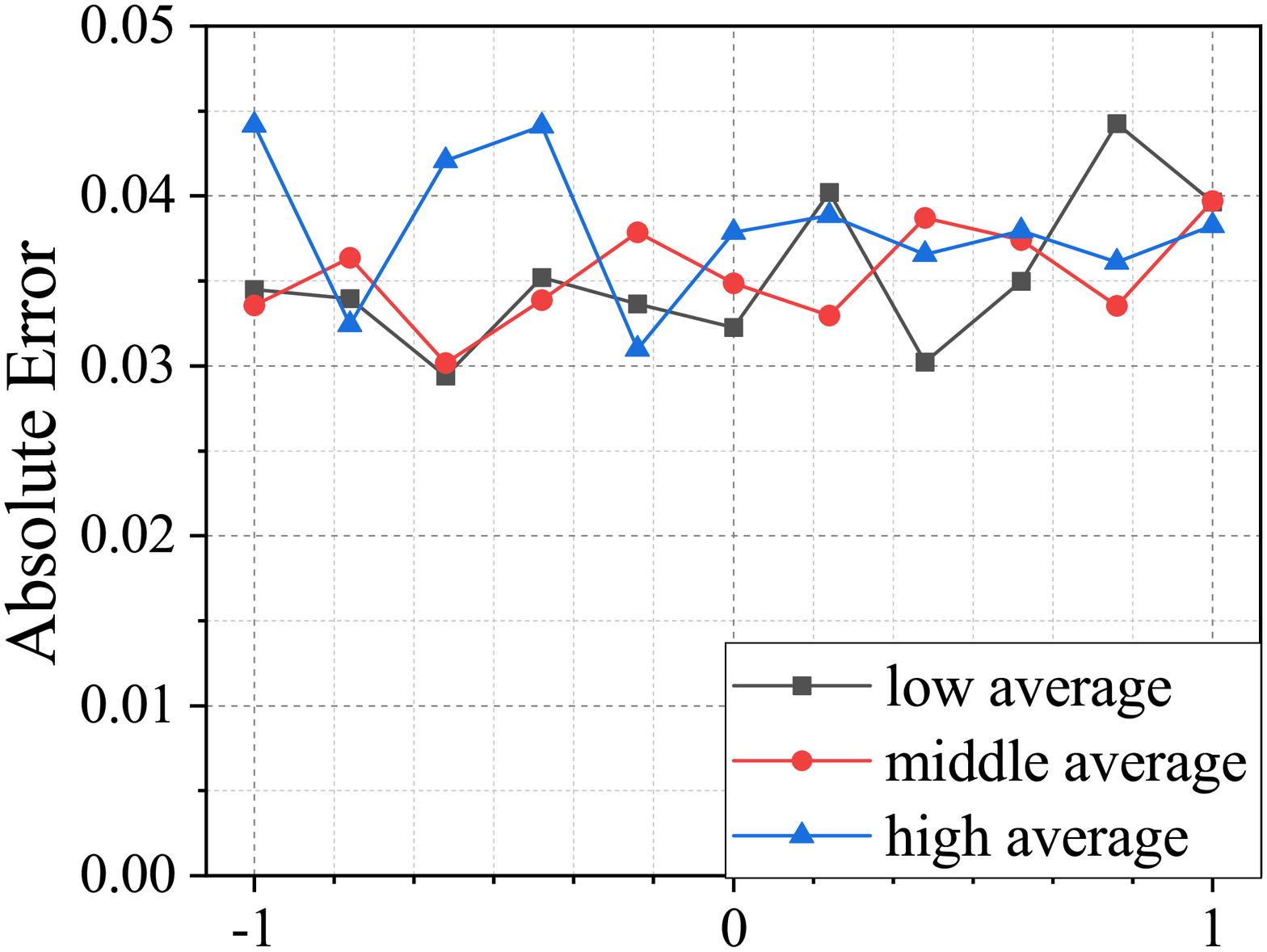} &
        \hspace{-12mm}\includegraphics[width=0.3\textwidth]{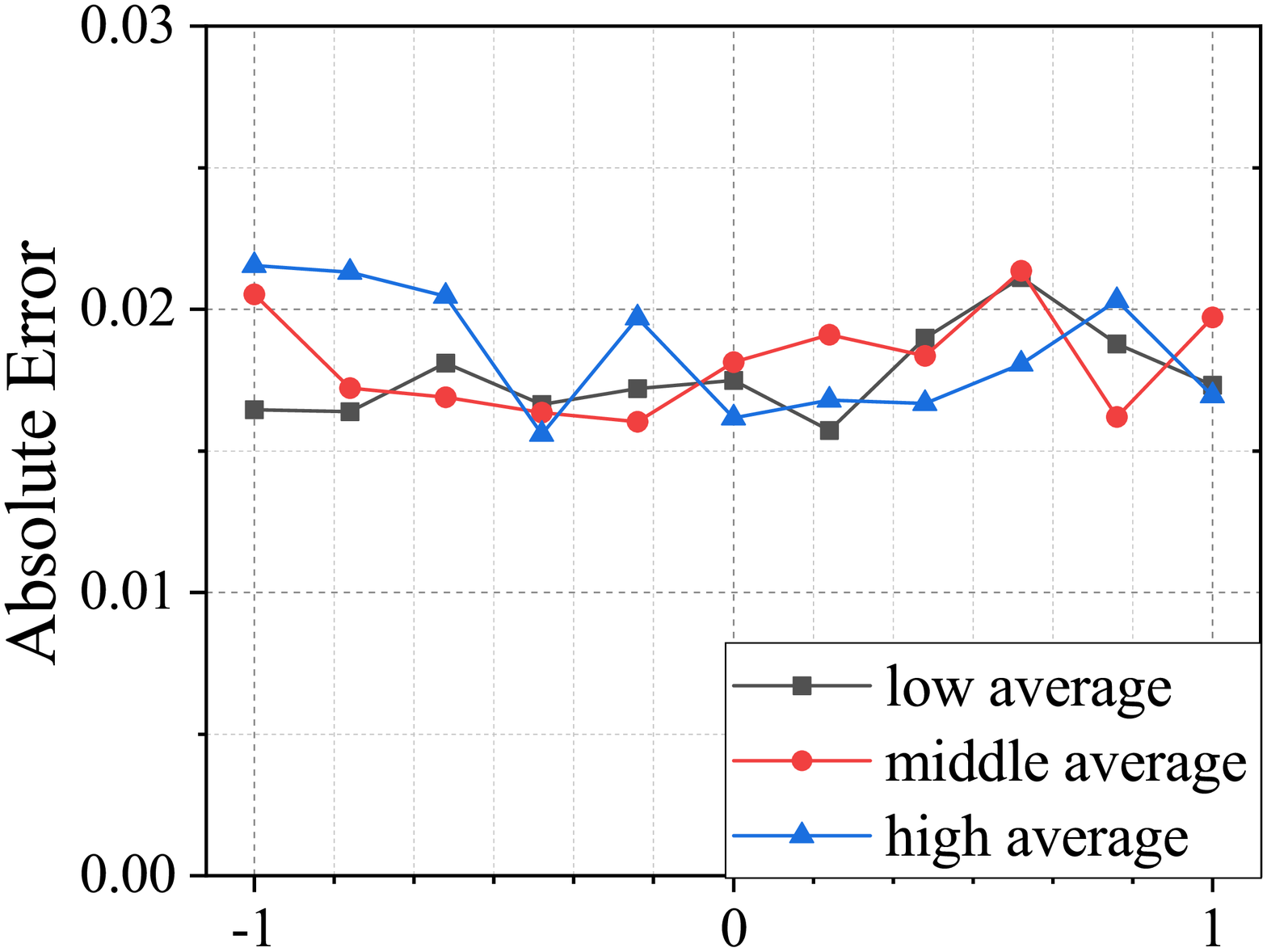}
        \\
        \hspace{-10mm} (a) epsilon = 0.5 & 
        \hspace{-12mm} (b) epsilon = 1 
    \end{tabular}
    \caption{\textit{F2M} under different default values.}
    \vspace{-8pt}
    \label{fig:f2m_default_value}
    \end{figure}

    Figure~\ref{fig:f2m_default_value} compares \textit{F2M} mechanisms with respect to different default values of $\overline{v}$. We observe that the performance of \textit{F2M} mechanism does not fluctuate when $\overline{v}$ changes, which reflects that the noise introduced by discretization is negligible compared to that by the randomized response.

\subsection{Conditional analysis}
    
    \begin{figure}[t]
    \centering
    \footnotesize
    \begin{tabular}{cc}\footnotesize
        \vspace{-10pt}
        \hspace{-8mm}\includegraphics[width=0.3\textwidth]{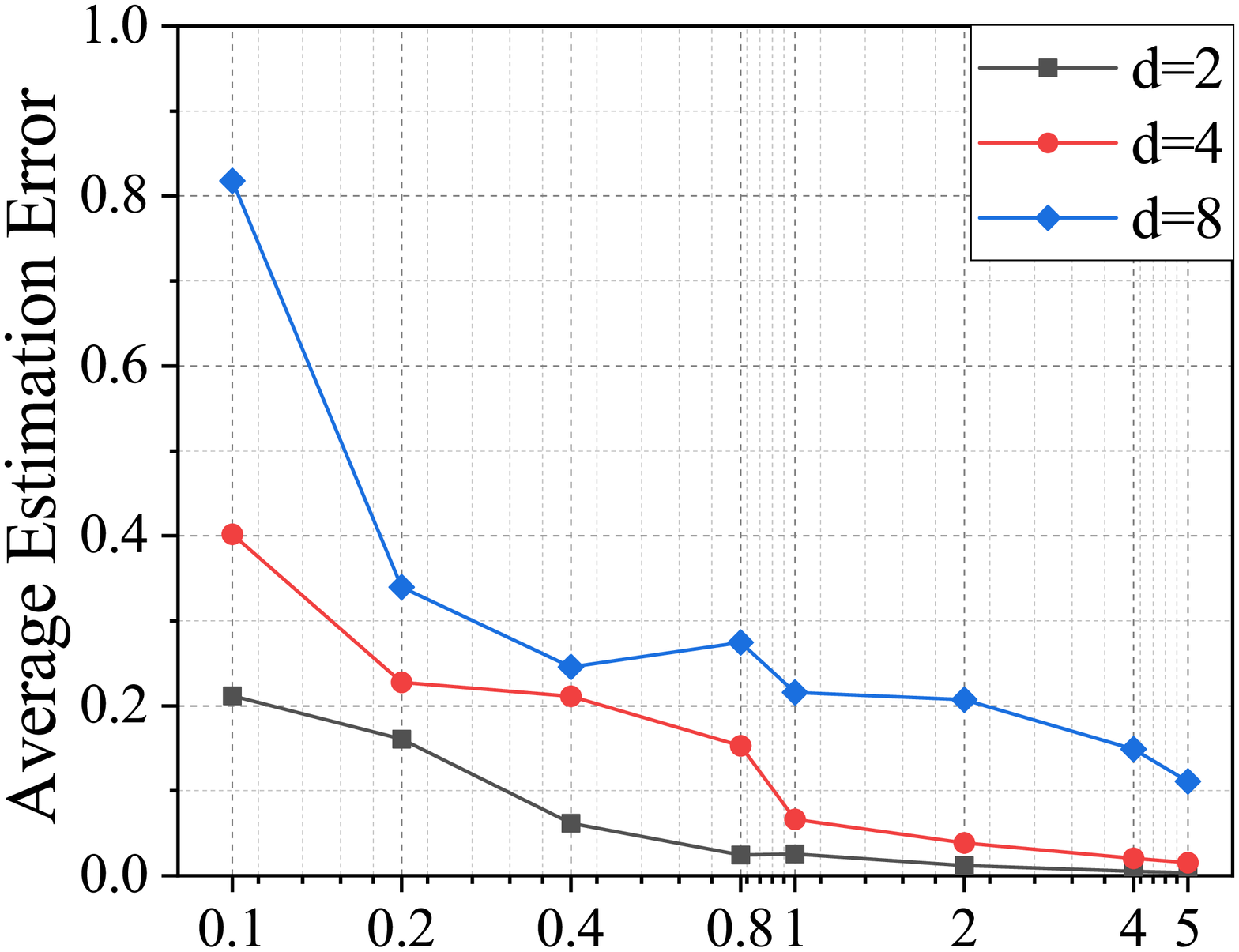}&
        \hspace{-12mm}\includegraphics[width=0.3\textwidth]{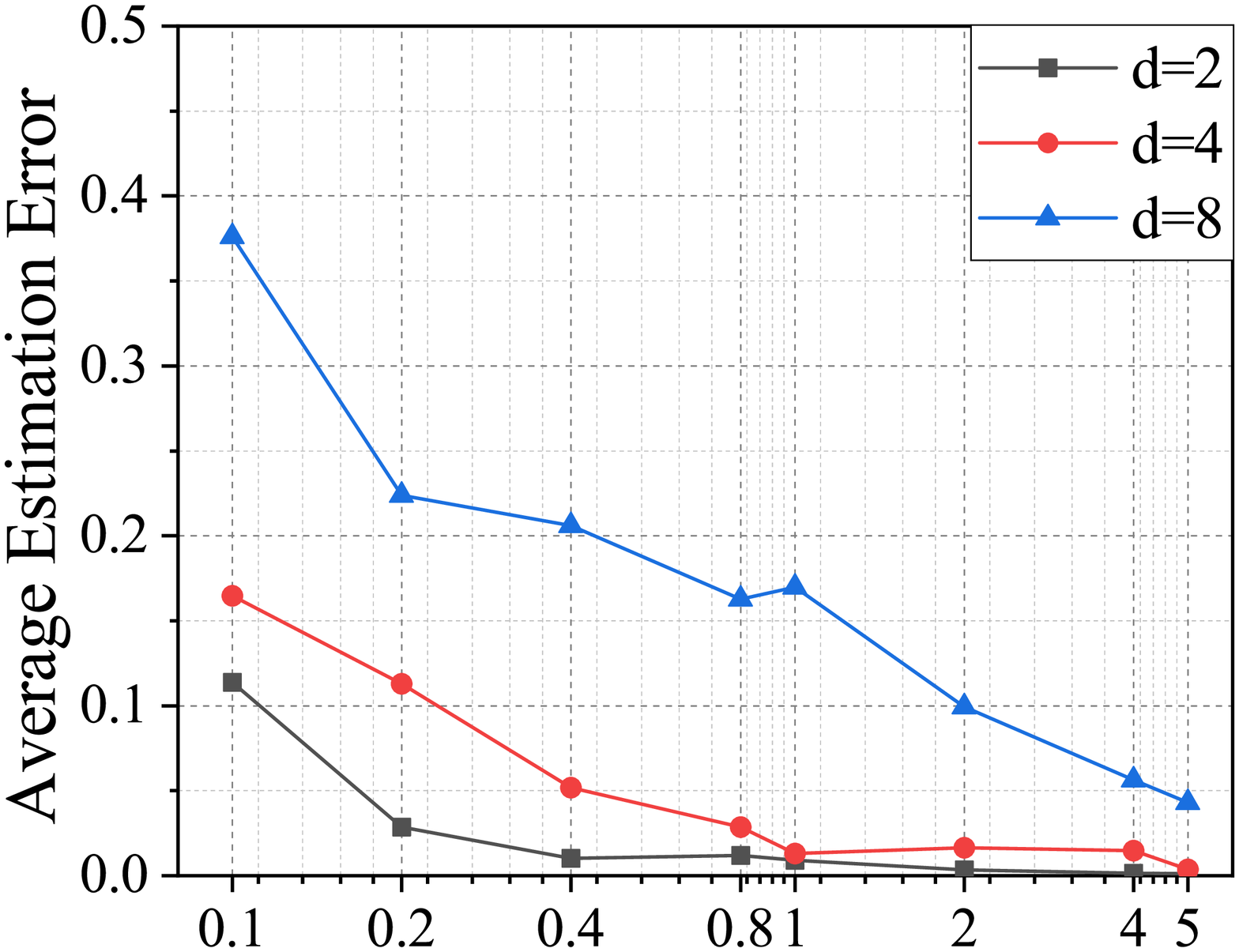}
        \\
        \hspace{-10mm} (a) N=$10^5$ & 
        \hspace{-12mm} (b) N=$10^6$
    \end{tabular}
    \caption{2-way conditional frequency estimation}
    \label{fig: conditional frequency}
    \end{figure}
    
    \begin{figure}[t]
    \vspace{-10pt}
    \centering
    \footnotesize
    \begin{tabular}{cc}\footnotesize
        \vspace{-10pt}
        \hspace{-8mm}\includegraphics[width=0.3\textwidth]{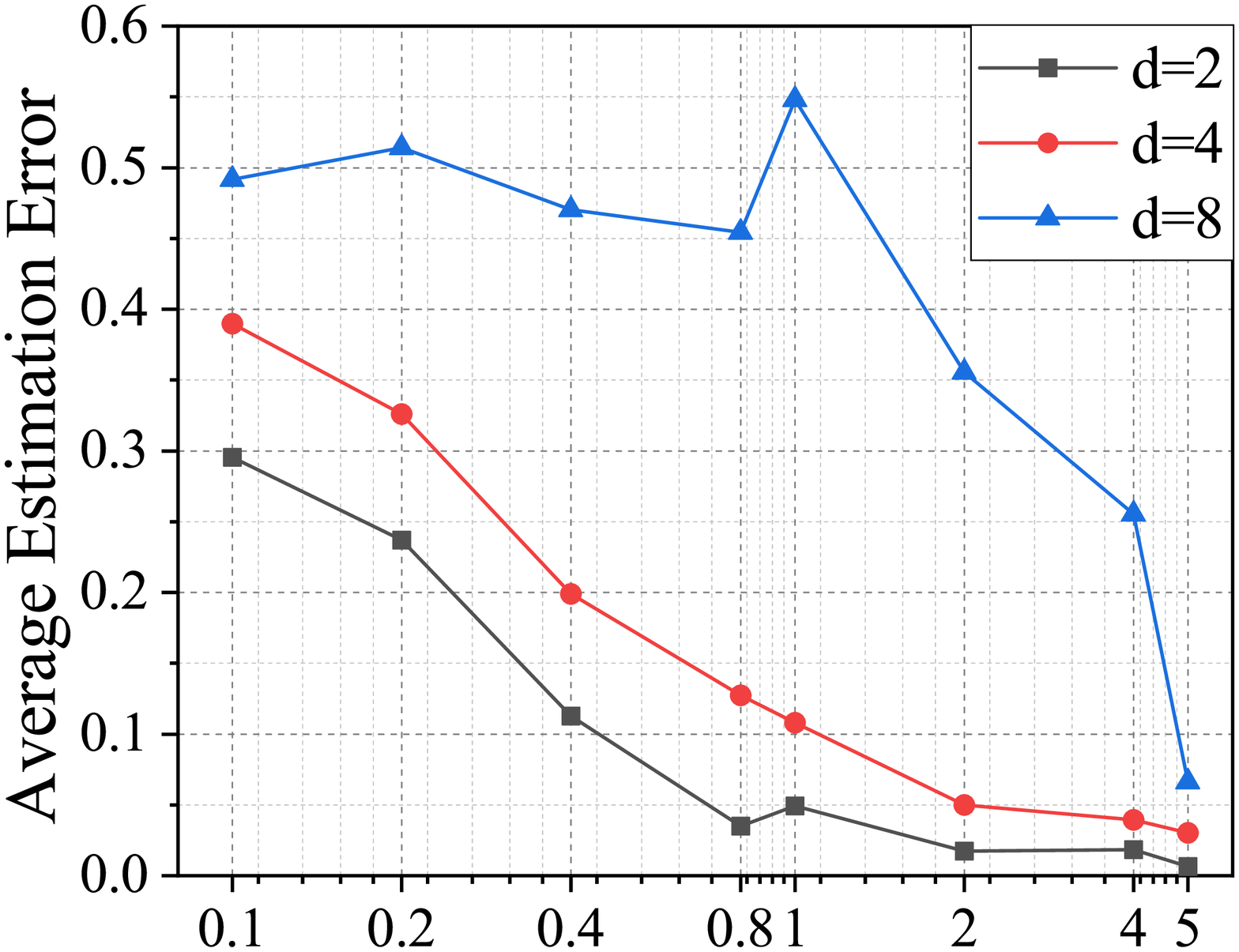}&
        \hspace{-12mm}\includegraphics[width=0.3\textwidth]{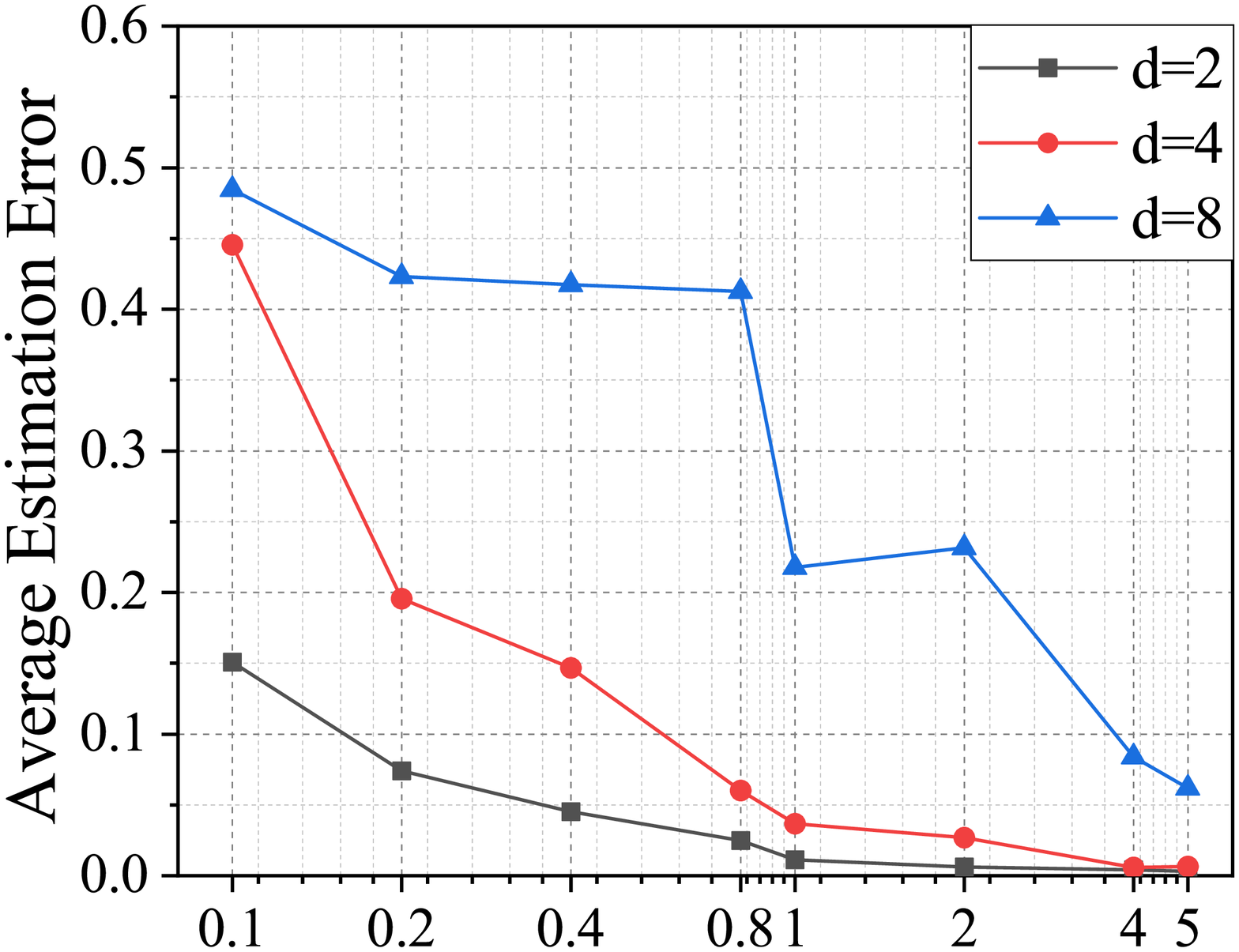}
        \\
        \hspace{-10mm} (a) N=$10^5$ & 
        \hspace{-12mm} (b) N=$10^6$
    \end{tabular}
    \caption{2-way conditional mean estimation}
    \vspace{-5pt}
    \label{fig: conditional mean}
    \end{figure}
    
    For the efficiency consideration, we only test the 2-way conditional analysis over $d \in \{2, 4, 8\}$ (with 20 observations under each configuration). We use datasets with $f_k=0.8$ in the low average case with $10^5$ and $10^6$ users. Figure~\ref{fig: conditional frequency} and ~\ref{fig: conditional mean} compare privately and non-privately computed conditional result of frequency and mean. We first figure out that the error of conditional mean estimation is lower to that of conditional mean estimation. We think this is because the error of frequency is involved in the mean estimation, as we analyzed in 1-Way frequency and mean estimation. We also observe that with the increase of dimensions, the estimating error increases, because the encoded space becomes huge ($\mathcal{O}(d^3)$). Thus, the conditional analysis has limitations on high-dimensional data currently.
    
\section{Conclusion and Future work}
\label{sec: conclusion}

    In this paper, we propose a series of locally differentially private mechanisms for frequency and mean estimation of key-value data. Based on the previous work of \textit{PrivKV}, we first propose a decoding mechanism for the data aggregator. Moreover, we combine several state-of-art LDP methods to improve the performance of frequency and mean estimation in the local settings. Theoretical analysis and empirical experiments validate the effectiveness and robustness of our proposed mechanisms. Beyond that, we introduce the notion of conditional analysis in key-value data analysis that allows the aggregator to learn the correlation between keys and corresponding values.
    
    The first part of work in the to-do list is to achieve an unbiased estimator for the mean. In this paper, we achieve low estimation error by unbiased estimation of the number of different key-value states after discretization. This leads to biased mean estimation. We will further show that we can achieve an unbiased estimator with the use of iteration. Besides that, to support conditional analysis in key-value data, we encode all of a user's data with one hot encoding mechanism. This takes cost in both communication and computation. Graham~\textit{et al.}~\cite{cormode2018marginal} use the Hadamard transform as evaluating a Hadamard entry is practically faster~\cite{bassily2017practical}. As our next move, we intend to improve the efficiency by the Hadamard transformation and improve accuracy by using an optimal encoding that achieves lower variance in the conditional analysis of key-value data.

\bibliographystyle{IEEEtran}
\bibliography{IEEEquote}

\appendix
\subsection{Error bound for \textit{F2M}}
\label{appendix: f2m}
    
    For the randomized response, assume there are $x$ records with value 1 of $N$ records. For the aggregator, after receiving $N$ records with $X$ records being 1, the estimated $x^*$ can be adjusted by:
    
    \begin{equation}
        x^* = X \cdot \frac{e^\epsilon+1}{e^\epsilon-1} - \frac{N}{e^\epsilon-1}.
    \end{equation}
    
    And we have $\mathbb{E}[x^*] = x$. According to the Chernoff--Hoeffding bound to independent $\{0,1\}$ random variables, for all $t>0$, we have $\Pr[ |X^* - X| \ge t ] \le 2e^{-\frac{2t^2}{N}}. $
    
    
    Setting $r = t\cdot \frac{e^\epsilon+1}{e^\epsilon-1}$ and $\delta = 2\cdot e^{-\frac{2r^2}{N} \cdot (\frac{e^\epsilon-1}{e^\epsilon+1})^2}$, we obtain
    
    \begin{equation}
        \Pr[|x - x^*| \ge r] \le \delta. \nonumber
    \end{equation}
    
    For the frequency, we have $x = N \cdot f_k$. And for the estimation of $m_{all}$, we have $x = N \cdot \frac{m_{all}+1}{2}$. Thus the error of $f$ is bounded by $r / N$ and the error of $m_{all}$ is bounded by $2r / N$. Then for the mean estimation, considering $N$ is big enough:
    
    
    \begin{align}
        | m_k^* - m_k| &= \left|\frac{m^*_{all} - (1-f_k^*)\cdot \overline{v}}{f_k^*} - \frac{m_{all} - (1-f_k)\cdot \overline{v}}{f_k} \right| \nonumber\\
        &= \left|\frac{f(m^*-m) + (f_k-f_k^*)(m-\overline{v})}{f\cdot f_k^*}\right| \nonumber\\
        &\le \left | \frac{2r/N}{f_k-r/N} + \frac{r/N}{f_k(f_k-r/N)}(m-\overline{v}) \right| \nonumber\\
        &= \frac{2r f_k+r(m-\overline{v})}{f_k(N f_k-r)} \le \frac{2r(f_k+1)}{f_k(N f_k-r)}. \nonumber
    \end{align}
    
    Thus, with probability at least $(1-\delta)^2$, we can assure:
    
    \begin{equation}
        |m_k^* - m_k| \le \frac{2(f_k+1)(e^\epsilon+1)\sqrt{\ln{(2/\delta)}}}{\sqrt{2N}f_k^2 (e^\epsilon-1) - f_k(e^\epsilon-1)\sqrt{\ln{(2/\delta)}}}. \nonumber
    \end{equation}

\end{document}